\documentclass[a4paper,11pt]{article}

\pdfoutput=1
\usepackage{jcappub}
\usepackage[T1]{fontenc}

\usepackage{lipsum}
\usepackage[margin = 1 in]{geometry}
\usepackage{amsmath,amsfonts,amssymb}
\usepackage{mathtools,physics,bm}
\usepackage{graphicx}
\usepackage{float}
\usepackage{empheq}
\usepackage{cuted}
\usepackage{indentfirst}
\usepackage{framed}
\usepackage{subfigure}
\usepackage[symbol]{footmisc}
\usepackage[export]{adjustbox}[2011/08/13]
\usepackage{comment}
\usepackage{changepage}

\usepackage{xcolor}

\newcommand{\PR}{\prime}
\newcommand{\MC}{\mathcal}
\newcommand{\lap}{\nabla^2}

\usepackage{amsthm}

\numberwithin{equation}{section}

\usepackage[normalem]{ulem}
\title{\boldmath Cosmological Constraints on 4-Dimensional Einstein-Gauss-Bonnet Gravity}

\author[a]{C. M. A. Zanoletti,}
\author[b,c]{B. R. Hull,}
\author[a]{C. D. Leonard,}
\author[c,d]{and R. B. Mann}

\affiliation[a]{\textit{School of Mathematics, Statistics and Physics, Newcastle University,\\ Herschel Building, Newcastle upon Tyne, NE1 7RU, United Kingdom}}
\affiliation[b]{\textit{Department of Applied Mathematics, University of Waterloo,\\
200 University Ave West, Waterloo, Ontario, N2L 3G1, Canada}}
\affiliation[c]{\textit{Perimeter Institute for Theoretical Physics,\\
31 Caroline Street North, Waterloo, Ontario, N2L 2Y5, Canada}}
\affiliation[d]{\textit{Department of Physics and Astronomy, University of Waterloo,\\
200 University Ave West, Waterloo, Ontario, N2L 3G1, Canada}}

\emailAdd{c.m.a.zanoletti2@newcastle.ac.uk}
\emailAdd{b2hull@uwaterloo.ca}

\abstract{4-Dimensional Einstein-Gauss-Bonnet (4DEGB) gravity has garnered significant attention in the last few years as a phenomenological competitor
     to general relativity. We consider the theoretical and observational  implications of this theory in both the early and late universe,  (re-)deriving background and perturbation equations and constraining its characteristic parameters with data from cosmological probes. 
     Our investigation surpasses the scope of previous studies by incorporating non-flat spatial sections. We explore consequences of 4DEGB on the sound and particle horizons in the very early universe, and demonstrate that 4DEGB can provide an independent solution to the horizon problem for some values of its characteristic parameter $\alpha$. Finally, we  constrain an unexplored regime of this theory in the limit of small coupling $\alpha$ (empirically supported in the post-Big Bang Nucleosynthesis era by prior constraints). This version of 4DEGB includes a geometric term that resembles dark radiation at the background level, but whose influence on the perturbed equations is qualitatively distinct from that of standard forms of dark radiation. In this limit, only one beyond-$\Lambda$CDM degree of freedom persists, which we denote as $\tilde{\alpha}_C$. Our analysis yields the estimate $\tilde{\alpha}_C = (-9 \pm 6) \times 10^{-6}$ thereby providing a new constraint of a previously untested sector of 4DEGB. 
     \newline
     \newline
     {\bfseries\large\sffamily{Keywords.}} 4-Dimensional Einstein-Gauss-Bonnet gravity, modified gravity, scalar-tensor theory, parameter estimation, horizon problem.}

\begin{document}
\maketitle
\flushbottom

\section{Introduction} \label{intro}

Cosmology aims to unravel the profound mysteries surrounding dark matter, dark energy and the early stages of cosmic evolution.
Although  general relativity (GR) has for over  100 years 
enjoyed immense empirical success \cite{Will_2014},  cosmological tensions (such as those in the $H_0$ and $S_8$ parameters \cite{Di_Valentino_2021}) hint towards the possible need for new physics to explain modern cosmological data in this era of precision cosmology. The prediction of singularities is also a troublesome consequence (and some would argue an impossibility) of GR.

A multitude of proposals have been put forth to address these unresolved issues in cosmology \cite{hu2023hubble}. One method that has garnered the attention of theorists, which will be the subject of this paper, is modified gravity. Stress testing GR by constraining alternative gravity theories with modern observations is an important step forward in confirming or rejecting its validity. Modified gravity has the ability to not only affect cosmology but gravitational phenomena as a whole, giving it an important role in the quest for a quantum theory of gravity. While this approach can be coupled with modifications to canonical particle theory, in this study we assume the Standard Model (SM) of particle physics.

Extensions of general relativity or alternative theories of gravity have been around since Einstein's discovery in 1915. This is a vast 
subject, and as such the reader is referred to \cite{Clifton_2012} for a concise and thorough review. One important result for the study of gravitational theory is Lovelock's theorem \cite{Lovelock:1971yv,Lovelock:1972vz}, which states that Einstein's field equations are the unique 4-dimensional 2$^{\textrm{nd}}$ order equations of motion for gravity that come from an action principle that solely relies on the metric. To deviate from general relativity one must therefore make modifications to the Einstein-Hilbert action \eqref{EHAction}
\begin{equation} \label{EHAction}
    S=\frac{1}{16 \pi G} \int \sqrt{-g}R \dd^4 x+\int \mathcal{L}_m\left(g_{\mu \nu}, \psi\right) \dd^4 x
\end{equation}
in a non-trivial way. A common procedure is to allow for external fields coupled to the curvature, as well as the addition of non-linear curvature terms in \eqref{EHAction}. Of interest here is the case of Gauss-Bonnet gravity, where one allows the unique set of quadratic invariants of the Riemann tensor \cite{Lanczos:1938sf,Lovelock:1971yv}
\begin{equation}\label{GBinv}
   \MC{G}=R^2-4 R_\nu^\mu R_\mu^\nu+R_{\rho \lambda}^{\mu \nu} R_{\mu \nu}^{\rho \lambda} 
\end{equation}
to be included in the action \eqref{EHAction}. 
Although the effects of Gauss-Bonnet gravity have been extensively studied in higher dimensional space-times, this invariant in 4 dimensions is a total derivative term, and as such does not contribute to spacetime dynamics. While it was initially reported in \cite{Glavan_2020} that a \textit{pure} Einstein-Gauss-Bonnet  theory of gravity
in 4 dimensions had been discovered (which we shall refer to as the Glavin-Lin (GL) formulation), it was established that there is no such theory \cite{Gurses:2020ofy,Ai:2020peo,Shu:2020cjw}; see \cite{Clifton_2020} for a  discussion on the formalism and shortcomings of \cite{Glavan_2020}.

These developments prompted others to consider how one would go about constructing a 4-dimensional theory of gravity that contains the Gauss-Bonnet term whilst retaining second order equations of motion and locality, and at the same time avoiding so-called ghosts  (theoretical instabilities leading to unphysical consequences). It was shown that this was possible by allowing the action to contain an additional scalar field, resulting in
 \cite{Robie4D,Fernandes:2020nbq}
\begin{equation} \label{ouraction}
    S=\int d^4 x \sqrt{-g} \frac{1}{16 \pi G}\left[R-2 \Lambda+\alpha\left(\phi \mathcal{G}+4 G^{\alpha \beta} \partial_\alpha \phi \partial_\beta \phi-4(\partial \phi)^2 \square \phi+2\left((\partial \phi)^2\right)^2\right)\right] + S_{m}
\end{equation}
which we shall refer to as 4-dimensional Einstein-Gauss-Bonnet (4DEGB) gravity. This is no longer a metric theory, but instead is within the class of modified gravity theories referred to as scalar-tensor theories and more specifically the Horndeski class \cite{Horndeski:1974wa}. A generalization of this theory containing additional terms dependent on $\phi$ can be obtained from dimensional reduction of 
Gauss-Bonnet gravity
\cite{LuPang}.
While there have been some investigations in the field of scalar-Einstein-Gauss-Bonnet gravity (sEGB) in 4 dimensions \cite{nojiri2023propagation}, this theory is distinct from 4DEGB. The higher derivative terms in \eqref{ouraction} are necessary and unique to 4DEGB  and are not present in sEGB.

Here we investigate some of the implications of 4DEGB gravity for cosmological observations, contributing to the already extensive efforts \cite{Fernandes_2022} at constraining this theory. Specifically, we consider a largely untested sector of 4DEGB parameterized by  $C$, a free parameter that comes from solutions to the scalar field equations of motion %\eqref{BGSSOL} 
derived from the action \eqref{ouraction}, in addition to  the coupling constant $\alpha$.
Since the GL formulation 
is obtained by considering the
$D\to 4$ limit of various $D$-dimensional solutions, it
has no counterpart for the parameter $C$. 
 
Several papers \cite{Clifton_2020,Banerjee:2020yhu,Banerjee:2020dad,Tangphati:2021mvu,Tangphati:2021tcy,Pretel:2021czp,
Feng_2021,Haghani_2020,Wang_Mota,Charmousis_2022,aoki2021cosmology,Wang_2021,Ge_2020,Sengupta_2022} have investigated the cosmological, astrophysical and solar system implications of the GL formulation (and also, to a lesser extent, of 4DEGB theory). Tests from the Weak-Field limit provide the tightest constraints 
\begin{equation}\label{eq:constraints}
 \alpha < 10^{10}\; \textrm{m}^2 \implies \Tilde{\alpha} \lesssim 10^{-43}
\end{equation}
 from LAGEOS satellite observations \cite{Clifton_2020}, where we define the dimensionless parameter $\Tilde{\alpha}$ as $\Tilde{\alpha} = \alpha H_0^2$. This limit on $\Tilde{\alpha}$ from weak field tests is 16 orders of magnitude smaller than those from gravitational wave constraints (see for example equation (148) of \cite{Fernandes_2022} for GW170817 constraints).
Requiring that atomic nuclei not be shielded by a horizon imposes the strongest bound on negative $\alpha$ \cite{Charmousis_2022}, yielding
\begin{equation}\label{eq:priors_alpha}
    -10^{-83} \lesssim \Tilde{\alpha} \lesssim 10^{-43}.
\end{equation}
 
 The terms in the field equations coming from 4DEGB gravity will be of the same order as those from GR for $\Tilde{\alpha} \sim \mathcal{O}(1)$ or $\alpha \sim \mathcal{O}(10^{52} \textrm{ m}^2)$. This means that in 4DEGB the $\alpha$ parameter alone provides negligible corrections to Einstein gravity in all regimes except in the very early universe, extremely close to a black hole \cite{Fernandes_2022}, or inside highly compact objects \cite{MichaelSarah}.
 
 The equations describing the Weak-Field limit are recovered almost exclusively by applying the Post-Newtonian expansion, which produces equations with no counterpart for the parameter $C$. Consequently, the constraints \eqref{eq:constraints} on $\alpha$ from the Weak-Field limit are valid for both the GL formulation and  4DEGB. This means that the parameter $\alpha$ is highly constrained 
 in 4DEGB \cite{Clifton_2020}, while few conclusive constraints on the parameter $C$ exist (see Section \ref{sec:IntroCosmoConstr} for a short overview).

The remainder of our paper is structured as follows: in Section \ref{Sec:Theory} we describe the theoretical background and derive the field and perturbation equations for 4DEGB. In Section \ref{sec:VeryEarlyUniverse} we introduce the remaining effects of the parameter $\alpha$, which will still impact universe dynamics at very early times even when priors \eqref{eq:priors_alpha} are enforced. We show that the sound horizon in 4DEGB diverges in the limit of early time, requiring the explicit introduction of an early-time cut-off. We also demonstrate how 4DEGB can provide an alternative to inflation in solving the horizon problem for some values of the parameter $\alpha$.

In Section \ref{sec:IntroCosmoConstr} we introduce a useful limit that allows us, in some regimes, to parameterize 4DEGB through a single additional degree of freedom compared to GR (the parameter $\tilde{\alpha}_C \equiv \frac{\alpha C^2}{H_0^2}$). We show that, in the small-$\alpha$ limit at times substantially past nucleosynthesis, 4DEGB is described by a parameter that at the background level introduces a ``dark radiation'' term and, at the level of the perturbation equations, modifies structure growth. In Section \ref{sec:alphaC_constraints_background} we make use of this formulation to place novel constraints on this heretofore minimally tested sector of 4DEGB using cosmic microwave background (CMB) data, finding $\tilde{\alpha}_C \equiv \frac{\alpha C^2}{H_0^2}  =(-9 \pm 6) \times 10^{-6}$. Section \ref{perts_lateuniverse} provides a qualitative analysis of the perturbative behaviour of the theory by analysing trends in the 4DEGB linear matter power spectrum, produced by modifying a GR Boltzmann solver. We conclude in Section \ref{sec:conc}.
\section{Theory \& Solutions}\label{Sec:Theory}

\subsection{Theoretical Background}

The 4DEGB theory \eqref{ouraction} is obtained using an approach   similar to that of the $D \to 2$ limit of general relativity \cite{Robb2D}. In this method  \cite{Robie4D,Fernandes:2020nbq}  
 one begins with the action
 \begin{equation}
    S_D^{\mathrm{GB}}=
    \frac{1}{16 \pi G_D}\left[
   \int d^D x \sqrt{-g}(R -2\Lambda)
+
    \frac{\alpha}{D-4}\left(\int d^D x \sqrt{-\tilde{g}} \tilde{\mathcal{G}}-\int d^D x \sqrt{-g} \mathcal{G}\right) 
    \right]
    +S_m 
\end{equation}
with  no prior assumptions about the curvature of the $D$-dimensional space, where $\tilde{g}_{\mu \nu} = \exp(\phi) g_{\mu \nu}$ and
$\tilde{\mathcal{G}}$ is the Gauss-Bonnet term \eqref{GBinv} for the metric   $\tilde{g}_{\mu \nu}$; $S_m$ is the matter action. Writing $\tilde{g}_{\mu \nu}$ in terms of ${g}_{\mu \nu}$ and taking the limit $D \to 4$ yields the action \eqref{ouraction}.

An alternate method for the construction of a 4DEGB theory is the Kaluza-Klein like approach \cite{LuPang,Kobayashi:2020wqy}, which begins with the action 
\begin{equation}
    S_D= \frac{1}{16 \pi G_D}\left[\int d^D x \sqrt{-g}(R + \alpha \mathcal{G}) \right] +S_m
\end{equation}
assuming a spacetime
\begin{equation}
    d s_D^2=d s_4^2+e^{2 \phi} d \Sigma_{D-4}^2
\end{equation}
where the 
$(D-4)$-dimensional space with line element $d\Sigma_{D-4}^2$ is maximally symmetric and is parameterized by the curvature $\lambda$, and $\phi$ is a scalar field. Dimensional reduction to
$D=4$ yields
\begin{equation}\label{action}
    \begin{aligned}
    S=\int d^4 x \sqrt{-g}& \frac{1}{16 \pi G}\left[R-2 \Lambda+\alpha\left( \phi \mathcal{G}+4 G^{\alpha  \beta} \partial_\alpha \phi \partial_\beta \phi-4(\partial \phi)^2 \square \phi+2\left((\partial \phi)^2\right)^2\right)\right. \\
    &\left. -2 \lambda R e^{-2 \phi}-12 \lambda(\partial \phi)^2 e^{-2 \phi}-6 \lambda^2 e^{-4 \phi}\right] + S_{m}
    \end{aligned}
\end{equation}
with manifest dependence on the curvature parameter  $\lambda$.  

The variation of \eqref{action} for spacetime dimension D with respect to the metric yields the gravitational field equations \cite{Robie4D,LuPang}

{\small
\begin{equation} \label{fieldequations}
\begin{aligned}
& \mathcal{E}_{\mu \nu} \equiv \Lambda g_{\mu \nu}+G_{\mu \nu}+\alpha\left[\phi H_{\mu \nu}-2 R\left[\left(\nabla_\mu \phi\right)\left(\nabla_\nu \phi\right)+\nabla_\nu \nabla_\mu \phi\right]+8 R_{(\mu}^\sigma \nabla_{\nu)} \nabla_\sigma \phi+8 R_{(\mu}^\sigma\left(\nabla_{\nu)} \phi\right)\left(\nabla_\sigma \phi\right)\right. \\
& \quad-2 G_{\mu \nu}\left[(\nabla \phi)^2+2 \square \phi\right]-4\left[\left(\nabla_\mu \phi\right)\left(\nabla_\nu \phi\right)+\nabla_\nu \nabla_\mu \phi\right] \square \phi-\left[g_{\mu \nu}(\nabla \phi)^2-4\left(\nabla_\mu \phi\right)\left(\nabla_\nu \phi\right)\right]\left(\nabla^2\right)^2 \\
&+8\left(\nabla_{(\mu} \phi\right)\left(\nabla_{\nu)} \nabla_\sigma \phi\right) \nabla^\sigma \phi-4 g_{\mu \nu} R^{\sigma \rho}\left[\nabla_\sigma \nabla_\rho \phi+\left(\nabla_\sigma \phi\right)\left(\nabla_\rho \phi\right)\right]+2 g_{\mu \nu}(\square \phi)^2-2 g_{\mu \nu}\left(\nabla_\sigma \nabla_\rho \phi\right)\left(\nabla^\sigma \nabla^\rho \phi\right) \\
& \quad-4 g_{\mu \nu}\left(\nabla^\sigma \phi\right)\left(\nabla^\rho \phi\right)\left(\nabla_\sigma \nabla_\rho \phi\right)+4\left(\nabla_\sigma \nabla_\nu \phi\right)\left(\nabla^\sigma \nabla_\mu \phi\right)+4 R_{\mu \nu \sigma \rho}\left[\left(\nabla^\sigma \phi\right)\left(\nabla^\rho \phi\right)+\nabla^\rho \nabla^\sigma \phi\right] \\
&\left.\quad+3 \lambda^2 e^{-4 \phi} g_{\mu \nu}-2 \lambda e^{-2 \phi}\left(G_{\mu \nu}+2\left(\nabla_\mu \phi\right)\left(\nabla_\nu \phi\right)+2 \nabla_\nu \nabla_\mu \phi-2 g_{\mu \nu} \square \phi+g_{\mu \nu}(\nabla \phi)^2\right)\right] = 8 \pi G T_{\mu \nu}
\end{aligned}
\end{equation}
}
where
\begin{equation*}
H_{\mu \nu}=2\left[R R_{\mu \nu}-2 R_{\mu \alpha \nu \beta} R^{\alpha \beta}+R_{\mu \alpha \beta \sigma} R_\nu^{\alpha \beta \sigma}-2 R_{\mu \alpha} R_\nu^\alpha -\frac{1}{4} g_{\mu \nu}\left(R_{\alpha \beta \rho \sigma} R^{\alpha \beta \rho \sigma}-4 R_{\alpha \beta} R^{\alpha \beta}+R^2\right)\right]
\end{equation*}
which identically vanishes in $ D \leq 4$.
The variation of \eqref{ouraction} with respect to the scalar field yields
\begin{equation} \label{scalarfieldequation}
\begin{aligned}
\mathcal{E}_\phi \equiv & -\mathcal{G}+8 G^{\mu \nu} \nabla_\nu \nabla_\mu \phi+8 R^{\mu \nu} \nabla_\mu \phi \nabla_\nu \phi-8(\square \phi)^2+8(\nabla \phi)^2 \square \phi+16 \nabla^a \phi \nabla^\nu \phi \nabla_\nu \nabla_\mu \phi \\
& +8 \nabla_\nu \nabla_\mu \phi \nabla^\nu \nabla^\mu \phi-24 \lambda^2 e^{-4 \phi}-4 \lambda R e^{-2 \phi}+24 \lambda e^{-2 \phi}\left[(\nabla \phi)^2-\square \phi\right]=0.
\end{aligned}
\end{equation}
We are reserving Greek indices for spacetime indices whereas Latin indices will be used for spatial ones. 
If $\lambda=0$ then these field equations reduce to those obtained from the action \eqref{ouraction}.
One elegant property of this theory is that the trace of \eqref{fieldequations} yields
\begin{equation} \label{Geometricconstraint}
    T=4 \Lambda-R-\frac{\alpha}{2} \mathcal{G}
\end{equation}
upon using \eqref{scalarfieldequation},  
where $T$ represents the trace of the stress energy tensor. 

In this paper we shall endeavor to place cosmological constraints on solutions to action \eqref{ouraction}, for which   $\lambda=0$. 

\subsection{FLRW Solutions}

In this section we will discuss the familiar FLRW metric in the context of 4DEGB gravity. We will first discuss the background spacetime and the associated field equations with a perfect fluid stress energy tensor. We will then move onto the perturbed spacetime, looking at scalar perturbations only. We will treat the perturbations of the stress energy tensor to be those of an anisotropic fluid, again with only scalar perturbations. 

The cosmological background and scalar perturbation equations in 4DEGB gravity have been previously studied \cite{Clifton_2020,Haghani_2020}; in this section, we re-visit their construction  afresh. We present our derivation in detail, which leads in the case of scalar perturbations to a more complete set of equations than previously presented in the literature. Furthermore, we expand upon the existing literature by explicitly including the possibility of a non-zero spatial curvature in the FLRW metric, with the resulting equations presented in Appendix
\ref{Appendix:equation_of_motion_curvature}.

\subsubsection{Background spacetime}\label{backgroundsection}

We begin with the familiar FLRW line element in co-moving coordinates with arbitrary spatial curvature
\begin{equation} \label{FRWcomving}
    \dd s^2=-\dd t^2+a(t)^2\left(\frac{\dd r^2}{1-k r^2}+r^2\left(\dd \theta^2+\sin (\theta)^2 \dd \phi^2\right)\right)
\end{equation}
where we parameterize to $k=+1,-1,0$ respectively corresponding to positive, negative, or flat spatial curvature, with  $a(t)$ the scale factor. Isotropy and homogeneity assumptions imply $\phi = \bar{\phi}(t)$ from which we obtain
\begin{equation} \label{eq:bacckgroundscalar}
    (a \ddot{\bar{\phi}}+\dot{a} \dot{\bar{\phi}}-\ddot{a})\left(a^2 \dot{\bar{\phi}}^2-2 a \dot{a} \dot{\bar{\phi}}+\dot{a}^2+k\right)=0
\end{equation}
from \eqref{scalarfieldequation} using
\eqref{FRWcomving}. We have suppressed the time dependence on $\bar{\phi}(t)$ and $a(t)$ and the dot represents a derivative with respect to coordinate time $t$,
with overbars denoting background quantities. 
We will also make use of the standard Hubble parameter
\[
H \equiv \frac{\dot{a}}{a}.
\]
The solutions to \eqref{eq:bacckgroundscalar} are
\begin{subequations} \label{phisols}
\begin{align} 
\dot{\bar{\phi}} & = H \pm \frac{\sqrt{-k}}{a} \label{kphisol} \\
\dot{\bar{\phi}} & =H +  \frac{A}{a} \label{BGSSOL}
\end{align}
\end{subequations}
where the former comes from
the second term in \eqref{eq:bacckgroundscalar}; clearly $k\leq 0$ for a real solution.  The 
latter solution \eqref{BGSSOL} comes from the first term in \eqref{eq:bacckgroundscalar}, with $A$   an arbitrary constant.  This is the most general solution, and we will henceforth take it to be the solution for the background scalar field.

For the field equations \eqref{fieldequations} we will use the perfect fluid stress energy with components 
$$T^{\mu}_{\phantom{\mu}\nu}=\text{diag}(-\bar{\rho},\bar{p},\bar{p},\bar{p})
$$ 
where both $\bar{\rho}$ and $\bar{p}$ are understood to be functions of coordinate time $t$. Using this, the
$tt$
component of \eqref{fieldequations} yields
\begin{equation} \label{hamconst}
    \frac{\left(-\dot{\bar{\phi}}^4 a^3+4 \dot{\bar{\phi}}^3 \dot{a} a^2-2 \dot{\bar{\phi}}^2\left(3 \dot{a}^2+k\right) a+4 \dot{a} \dot{\bar{\phi}}\left(\dot{a}^2+k\right)\right) \alpha}{a^3}+\frac{\dot{a}^2+k}{a^2}=\frac{8 \pi G \bar{\rho}}{3}
\end{equation}
which through the use of \eqref{BGSSOL} and the Hubble parameter becomes
\begin{equation}\label{tthubbleBG}
\alpha\left(H^2-\frac{A^2}{a^2}\right)
    \left(H^2+\frac{A^2+2k}{a^2}\right)
    +H^2+\frac{k}{a^2}
    =\frac{8 \pi G \bar{\rho}}{3}.
\end{equation}
Without loss of generality we can set $A^2=-k+C$, obtaining
\begin{equation} \label{constraintfried}
    \left(H^2+\frac{k}{a^2}\right)^2 \alpha+H^2+\frac{k}{a^2}=\frac{8 \pi G \bar{\rho}}{3} + \frac{\alpha C^2}{a^4}
\end{equation}
where the arbitrary constant $C$ now acts as a geometrical ``dark radiation''  parameter \cite{Clifton_2020}.  If $C=0$ we recover the generalization of the Friedmann equations obtained in \cite{Glavan_2020}. 
 
The other generalized Friedmann equation can be obtained from
  any one of the diagonal terms of \eqref{fieldequations}, yielding 
 {\small 
\begin{equation}
   \frac{\left(4\left(H^2 a^4+k a^2\right) \dot{H}+3 H^4 a^4+2 k H^2 a^2+A^2\left(A^2+2 k\right)\right) \alpha}{a^2}+3 H^2 a^2+2 \dot{H} a^2+k = - 8 \pi G a^2 \bar{p}
\end{equation}}
or, setting $A^2=-k+C$,  

\begin{equation} \label{evolfried}
    \frac{\left(H^2 a^2+k\right)\left(3 H^2 a^2+4 \dot{H} a^2-k\right) \alpha}{a^2}+3 H^2 a^2+2 \dot{H} a^2+k = - 8 \pi G a^2 \bar{p} -\frac{\alpha C^2}{a^2}.
\end{equation}

The conservation of stress energy yields
\begin{equation}
\begin{aligned}
    \nabla_\mu T^{\mu \nu}=0 \quad \rightarrow \quad & \dot{\bar{\rho}} a+3 \dot{a} \bar{\rho}+3 \dot{a} \bar{p} =0 \\
    & \dot{\bar{\rho}}=\frac{- 3 (\bar{\rho}+\bar{p}) \dot{a}}{a}
    \end{aligned}
\label{bgndcons}    
\end{equation}
and the system \eqref{constraintfried}, \eqref{evolfried}, and 
\eqref{bgndcons} comprises the background equations for 4DEGB. Note that  \eqref{evolfried} is a consequence of \eqref{constraintfried}  and 
\eqref{bgndcons}, as in general relativity.

Before we consider   perturbations about the background solution, it will be useful to consider these equations in conformal time, where  the 
line element \eqref{FRWcomving} is now written as 
\begin{equation} \label{FRWCON}
    \mathrm{d} s^2=a(\eta)^2\left(-\mathrm{d} \eta^2+\frac{\mathrm{d} r^2}{1-k r^2}+r^2\left(\mathrm{~d} \theta^2+\sin (\theta)^2 \mathrm{~d} \phi^2\right)\right)
\end{equation}
with
\begin{equation} \label{eq:conformal}
    \dd \eta =\frac{\mathrm{d} t}{a(t)}\; .
\end{equation}
Letting prime denote a derivative with respect to $\eta$, 
the Hubble parameter becomes
\begin{equation}
 \mathcal{H}(\eta) \equiv \frac{a^{\prime}(\eta)}{a(\eta)} \quad   \quad \Rightarrow \MC{H}(\eta) = a(t) H(t)
\end{equation}
with $\bar{\rho}$ and $\bar{p}$ now functions of $\eta$, along with the scalar field $\bar{\phi}$, with solution
\begin{equation} \label{eq:bgphiconformal}
\bar{\phi}^{\prime}= \mathcal{H} + A
\Rightarrow 
   \bar{\phi} = \ln(a) + A \eta + B \; .
\end{equation}
In conformal time the generalized Friedmann equations are now written as
\begin{equation}\label{eq:boxed_background_eq}
    \begin{aligned}
        & \frac{\left( \MC{H}^2+k\right)^2 \alpha}{a^2}+ \MC{H}^2+ k = \frac{8 \pi G a^2 \bar{\rho} }{3} +\frac{ \alpha C^2}{a^2}\\
        & \frac{\left(\MC{H}^2+k\right)\left( 4 \MC{H}^{\PR} -\MC{H}^2-k\right) \alpha}{a^2}+\MC{H}^2+2 \MC{H}^{\PR}+k = - 8 \pi G a^2 \bar{p} -\frac{\alpha C^2}{a^2}  \\
        & \bar{\rho}^{\PR}=-\frac{3 a^{\PR}(\bar{p}+\bar{\rho})}{a}.
    \end{aligned}
\end{equation} 
 The use of a conformal or co-moving metric is a matter of taste. We shall
 employ  the former (unless otherwise noted) since $\bar{\phi}^{\prime}$ depends on the scale factor only when in terms of the
 Hubble parameter, as we see in
 \eqref{eq:bgphiconformal}.

\subsubsection{Perturbed Spacetime}\label{Theory:PerturbedSpacetime}

We will work in the context of the Conformal Newtonian  gauge, and only consider scalar perturbations. All 
leading order equations are  \textit{linear} order in the
perturbative quantities, which  depend on all spacetime coordinates.  The perturbed  version of the line element \eqref{FRWCON} is
\begin{equation} \label{pertmetric}
    \mathrm{d} s^2=a^2\left(-(1+2 \Phi) \mathrm{d} \eta^2+(1-2 \Psi)\left(\frac{\mathrm{d} r^2}{1-k r^2}+r^2\left(\mathrm{~d} \theta^2+\sin (\theta)^2 \mathrm{~d} \phi^2\right)\right)\right)
\end{equation}
where $\Phi$ and $\Psi$ are the metric scalar perturbations. We follow the convention of \cite{Fernandes_2022} in the labelling of $\Phi$ and $\Psi$. Perturbations of the stress energy \cite{Clifton_2020,Malik_2009} are given by $\delta \rho$ and $\delta p$ for the density and pressure respectively, as well as the velocity perturbation $v$ and  the anisotropic fluid perturbation $\pi_{ij}$ through the following
\cite{Clifton_2020,Malik_2009}
\begin{equation} \label{stresspert}
 \delta T_0^0=-\delta \rho, \quad \delta T_i^0=(1+w) \bar{\rho}\partial_i v, \quad \delta T_j^i=\delta_j^i \delta p+\pi_j^i
\end{equation}
where $\pi$ is defined via
\begin{equation}
    \pi_{ij}=\left(\nabla_i \partial_j-\frac{1}{3} \gamma_{ij}\lap \right) \Pi
\end{equation}
and where $\Pi$ is a scalar function representing the anisotropy, $\nabla_i$ represents a covariant derivative\footnote{In the absence of spatial curvature one may replace $\nabla_i$ with $\partial_i$ in $\pi_{ij}$, consistent with \cite{Clifton_2020,Malik_2009}. }, and $\lap$ is the Laplacian with respect to the spatial metric $\gamma_{ij}$. 

The relative energy density perturbation and speed of sound are respectively defined
as
\begin{equation}
    \boldsymbol{\delta} = \frac{\delta \rho}{\bar{\rho}} \qquad c_s^2=\frac{\delta p}{ \delta \rho}
\end{equation}
and  
\begin{equation}
    \phi \rightarrow \bar{\phi} + \delta \phi
\end{equation}
defines the perturbation of the  scalar field. 

We now consider the perturbed equations of motion, starting with the scalar field equation \eqref{scalarfieldequation}.
Its leading order correction is
\begin{equation} \label{firstorderscalarweqn}
    \alpha\left(A^2+k\right)\left(\lap(\delta \phi - \Phi)+ 3 A (\Psi^{\PR}+\Phi^{\PR}) - 3 (\Psi^{\PR \PR}+ \delta \phi^{\PR \PR} )\right)=0
\end{equation}
upon using \eqref{eq:bgphiconformal} and \eqref{pertmetric}. Note that it has no dependence on the scale factor $a$, which would not be the case - as stated above - if we were working in co-moving coordinates. Setting $A^2=-k+C$, this becomes
\begin{equation} \label{Cfirstorder}
    \alpha C \left(\nabla^2(\delta \phi-\Phi)+3 \sqrt{-k+C}\left(\Psi^{\prime}+\Phi^{\prime}\right)-3\left(\Psi^{\prime \prime}+\delta \phi^{\prime \prime}\right)\right)=0.
\end{equation}
Similar to the discussion above, when $C=0$ the equation is trivially satisfied and no information about the perturbation $\delta \phi$ is given. 

Henceforth we  focus on the situation in which $C \ne 0$ and $k=0$ (with the $k\ne0$ case presented in Appendix \ref{Appendix:equation_of_motion_curvature}).   
Turning now to   the gravitational field equations \eqref{fieldequations}, the 
 $\MC{E}_{\eta \eta} = 8 \pi G T_{\eta \eta} $ component is
 \begin{footnotesize}
\begin{equation} \label{eqn1}
    2\alpha A^2\left( 3 \Phi A^2 - 6 A(\Psi^{\PR}+ \delta \phi^{\PR}) - 2 \lap (\Psi +  \delta \phi) \right) +  \mathcal{A}(2 \lap \Psi - 6 \mathcal{H}\Psi^{\PR}) - 6 \alpha \Phi \mathcal{H}^4=8 \pi Ga^2(a^2 \delta \rho+2 a^2 \Phi \rho)
\end{equation}
\end{footnotesize}
at leading order, where $\delta T_{\eta \eta  } = a^2 \delta \rho+2 a^2 \Phi \bar{\rho} $ 
and we follow \citep{Clifton_2020} in defining
\begin{equation}
   \mathcal{A} \equiv 2 \alpha \mathcal{H}^2+a^2 \qquad 
   \mathcal{B} \equiv 2 \alpha \mathcal{H}^2-a^2
\end{equation}
for convenience.
From the  zeroth order equation  we identify
\[ \bar{\rho}=\frac{-3 A^4 \alpha+3 \alpha \MC{H}^4+3 a^2 \MC{H}^2}{8 a^4 \pi G}
\]
and inserting this
into \eqref{eqn1} we find
\begin{equation} \label{OUREQ1}
    4 \alpha A^2 \left( 3 A^2 \Phi - 3 
    A ( \Psi^{\PR} + \delta \phi ^{\PR}) - \lap( \Psi +  \delta \phi)\right) + 2 \mathcal{A} \left( \lap \Psi - 3 \mathcal{H}\Psi^{\PR} - 3 \Phi \mathcal{H}^2 \right) = 8 \pi G a^4 \delta \rho.
\end{equation}
 For any time-spatial component $\MC{E}_{\eta i}= 8 \pi G T_{\eta i}$ we obtain 
\begin{equation*}
    \left( 4\alpha A^2(A(\Phi - \delta \rho)-\Psi^{\PR}-\delta \rho^{\PR}) +2\mathcal{A}\Psi^{\PR}+2\mathcal{H} \mathbf{\MC{A}}\Phi \right),_i=-8 \pi G a^4 (1+w)\rho \partial_i v
\end{equation*}
to leading order, which  readily integrates to
\begin{equation}\label{OUREQ2}
    4\alpha A^2(A(\Phi -\delta \rho)-\Psi^{\PR}-\delta \rho^{\PR}) +2\mathcal{A}\Psi^{\PR}+2\mathcal{H} \mathbf{\MC{A}}\Phi = -8 \pi G a^4 (1+w)\rho v.
\end{equation}
For the spatial-spatial components of $\mathcal{E}_{ij}= 8 \pi G T_{ij}$ where $i \ne j$  we have
\begin{equation*}
    \left( 2 \alpha (A^2(\Psi+\Phi)+ 2 \mathcal{H}^{\PR} \Psi) - \mathcal{B} \Psi - \mathcal{A} \Phi \right),_{ij}=8 \pi G a^4 \Pi ,_{ij}
\end{equation*}
 to leading order, which again easily integrates into
\begin{equation}\label{OUREQ3}
    2 \alpha\left(A^2(\Psi+\Phi)+2 \mathcal{H}^{\prime} \Psi\right)-\mathcal{B} \Psi-\mathcal{A} \Phi=8 \pi G a^4 \Pi
\end{equation}
where we have set any integration functions of $\eta$ to zero.  This can be done without any loss of generality as the constraints must hold for all time;  as such we can tune them to any value we wish.  We note that for time-spatial and mixed spatial components there are no zeroth order counterparts.

Lastly we consider the diagonal components, $\MC{E}_{ii}=8 \pi G T_{ii}$. Since we lose homogeneity and isotropy at leading order, the equations are no longer equivalent for $i=x,y,z$, unlike the zeroth order case. Due to this feature we take the trace, $\MC{E}^{i}_{i}=8 \pi G T^{i}_{i}$ which to leading order can be written as \begin{equation}
    \delta G_{xx}+\delta G_{yy}+\delta G_{zz}=8 \pi G (3 a^2 ( \delta p - 2\Psi \bar{p})).
\end{equation}
Now isolating
\begin{equation} \label{presszero}
    \bar{p}=\frac{-A^4 \alpha+\alpha H^4-4 \dot{H} H^2 \alpha-H^2 a^2-2 \dot{H} a^2}{8 a^4 \pi G}
\end{equation}
from any zeroth order diagonal equation and inserting into $\delta G_{xx}+\delta G_{yy}+\delta G_{zz}=8 \pi G (3 a^2 ( \delta p - 2\Psi \bar{p}))$
we get the final expression
{\small
\begin{equation}\label{OUREQ4}
    \begin{aligned}
        & 2\alpha \left( 3 A^2  \Phi - 3A( \delta \phi^{\PR} - \Phi^{\PR}) - \lap( \Phi + \Psi) - 3(\delta \phi^{\PR \PR} + \Psi^{\PR \PR}) \right) + \MC{A}( 3 \MC{H} \Phi^{\PR} + \lap \Phi + 3 \Psi^{\PR \PR}) + \MC{B} \lap \Psi \\
        & + 6 \MC{H}^{\PR} \MC{C} \Phi + 6 \MC{H}\MC{D} \Psi^{\PR} - 3 \MC{H}^2 \MC{B} \Phi - 4 \MC{H}^{\PR} \alpha \lap \Psi = 12 \pi G a^4 \delta p
    \end{aligned}
\end{equation}
where $\MC{C} = 4 \alpha \MC{H}^2 + a^2$ and $\MC{D}= 2 \MC{H}^{\PR} \alpha + a^2$. 
We note here a subtle point   regarding the sum of the diagonal components, which yields three equations that  are not identical at leading order. However making use of \eqref{OUREQ3} renders these equations   equivalent and hence the sum may be taken without any loss of generality. This subtle feature sometimes gets taken for granted, even when discussing General Relativity. 

At this point we have 7 unknown scalar functions -- $\Phi,\Psi,\delta \phi,v,\delta \rho ,\delta p$ and $\Pi$ -- and 
 5 equations: \eqref{Cfirstorder} from the  scalar field equation, and  \eqref{OUREQ1},\eqref{OUREQ2},\eqref{OUREQ3}, and \eqref{OUREQ4}  from the gravitational field equations \eqref{fieldequations}.
The remaining two equations come from the conservation of stress-energy $\nabla_{\mu} T^{\mu \nu}=0$. At zeroth order we retrieve only one equation, the cosmological fluid equation
\begin{equation} \label{zerostresssol}
    \bar{\rho}^{\prime}=-3 \MC{H}(\bar{p}+\bar{\rho})
\end{equation}
obtained  above. The first order $\nabla_{\mu} T^{\mu \eta}$ equation is
\begin{equation} \label{leadstress1}
    \bar{\rho} \left[(1+w)\lap v - 3 (\Psi^{\PR}+2\MC{H} \Phi) \right] - 3 \Psi^{\PR} \bar{p}+ \delta \rho^{\PR} - 2 ( 3\MC{H} \bar{p} + \bar{\rho}^{\PR}) \Phi + 3 \MC{H}( \delta p + \delta \rho)=0
\end{equation}
and after inserting the zeroth order solution \eqref{zerostresssol} this becomes
\begin{equation} \label{zerostresssol2}
    \delta \rho^{\PR} = (1+w)(3 \Psi^{\PR}-\lap v) \bar{\rho}- 3 \MC{H}(\delta \rho +\delta p).
\end{equation}
This can be rewritten as
\begin{equation}
    \boldsymbol{\delta}^{\PR} = 3 \mathcal{H} \boldsymbol{\delta}\left(w-c_s{}^2\right) +(1+w)\left(3 \Psi^{\prime}-\lap v \right)
\end{equation}
using  $ \boldsymbol{\delta} \bar{\rho} = \delta \rho \Rightarrow \delta \rho^{\PR}= \boldsymbol{\delta}^{\PR} \bar{\rho}+\boldsymbol{\delta} \bar{\rho}^{\PR}$.
Lastly, any $\nabla_{\mu} T^{\mu i}=0$ component - they are all equivalent - in conjunction with \eqref{zerostresssol} yields
\begin{equation}
    \partial_i (2  \lap \Pi + 3 \delta p + 3(1+w) \bar{\rho}[ (1-3w)v+v^{\PR}+\Phi])=0
\end{equation}
whose divergence is
\begin{equation}
    2 \nabla^{4} \Pi + 3 \lap \delta p + 3(1+w)\bar{\rho}[(1-3w)\lap v + \lap v^{\PR}+ \lap \Phi]=0,
\end{equation}
a form that will prove useful later on.

Upon setting  $A^2=C$,   the full set of equations for $k=0$ is
\begin{equation}\label{eq:boxed_pert_eq}
\mathclap{
    \begin{aligned}
        & \alpha C \left(  \nabla^2(\delta \phi-\Phi)+3 \sqrt{C}\left(\Psi^{\prime}+\Phi^{\prime}\right)-3\left(\Psi^{\prime \prime}+\delta \phi^{\prime \prime}\right) \right) =0 \\
        & 4 \alpha C\left(3 C \Phi-3 \sqrt{C}\left(\Psi^{\prime}+\delta \phi^{\prime}\right)-\nabla^2(\Psi+\delta \phi)\right)+2 \mathcal{A}\left(\nabla^2 \Psi-3 \mathcal{H} \Psi^{\prime}-3 \Phi \mathcal{H}^2\right)=8 \pi G a^4 \delta \rho \\
        & 4 \alpha C\left(\sqrt{C}(\Phi-\delta \rho)-\Psi^{\prime}-\delta \rho^{\prime}\right)+2 \mathcal{A} \Psi^{\prime}+2 \mathcal{H} \mathcal{A} \Phi=-8 \pi G a^4(1+w) \bar{\rho} v \\
        & 2 \alpha\left(C(\Psi+\Phi)+2 \mathcal{H}^{\prime} \Psi\right)-\mathcal{B} \Psi-\mathcal{A} \Phi=8 \pi G a^4 \Pi \\
        & \begin{aligned}
& 2 \alpha\left(3 C \Phi-3 \sqrt{C}\left(\delta \phi^{\prime}-\Phi^{\prime}\right)-\nabla^2(\Phi+\Psi)-3\left(\delta \phi^{\prime \prime}+\Psi^{\prime \prime}\right)\right)+\mathcal{A}\left(3 \mathcal{H} \Phi^{\prime}+\nabla^2 \Phi+3 \Psi^{\prime \prime}\right) \\
& +\mathcal{B} \nabla^2 \Psi+6 \mathcal{H}^{\prime} \mathcal{C} \Phi+6 \mathcal{H} \mathcal{D} \Psi^{\prime}-3 \mathcal{H}^2 \mathcal{B} \Phi-4 \mathcal{H}^{\prime} \alpha \nabla^2 \Psi=12 \pi G a^4 \delta p
\end{aligned} \\
& \boldsymbol{\delta}^{\prime}=3 \mathcal{H} \boldsymbol{\delta}\left(w-\frac{\delta p}{\delta \rho}\right)+(1+w)\left(3 \Psi^{\prime}-\nabla^2 v\right) \\
& 2 \nabla^4 \Pi+3 \nabla^2 \delta p+3(1+w) \bar{\rho}\left[(1-3 w) \nabla^2 v+\nabla^2 v^{\prime}+\nabla^2 \Phi\right]=0.
    \end{aligned}
    }
\end{equation}
 The full set of equations with $k\neq 0$ are given in Appendix \ref{Appendix:equation_of_motion_curvature}.

In what follows, we will consider these equations from two perspectives. First, we will look at the full solutions at the background level outlined by equations \eqref{eq:boxed_background_eq},  which describe phenomena in the very early Universe. We shall then investigate an unexplored  
region of parameter space in which $\alpha$ is small but $C$ is unconstrained. Mathematically this is most easily studied by
taking $\alpha \rightarrow 0$ but keeping $\alpha C^2$  finite. 
This will have interesting implications for cosmological observables at CMB redshifts and later times.
\section{The very early universe}\label{sec:VeryEarlyUniverse}

When $a$ is small or $H$ is large the $\alpha H^4$ term in \eqref{constraintfried} will be non-negligible (even for small $\alpha$). Therefore, in the very early universe the parameter $\alpha$ will have an effect on background dynamics.  We will primarily explore the effects of $\alpha$ on the sound and particle horizons in the early universe and on the horizon problem.

\subsection{Diverging horizons close to the Big Bang}\label{Diverging horizons}

One striking feature of 4DEGB is its effect on the comoving sound horizon $r_s(z)$, where
$a=\frac{1}{1+z}$. This is defined as the comoving distance a sound wave could travel from the beginning of the universe in some time $t(z)$. The sound horizon at recombination is a fundamental scale shaped by early universe physics that leaves an imprint on the clustering of matter in the universe \cite{Alam_2021}.  The equation describing $r_s(z)$ is  
\begin{equation}\label{eq:SoundHorizonLCDM}
    r_s(z) = \int^{\infty}_z \frac{c_s(z')}{H(z')}\text{d}z' = c\int^{a}_{0} \frac{1}{H(a')a'^2\sqrt{3(1+R)}}\text{d}a'
\end{equation}
where $c_s(z) = \frac{c}{\sqrt{3(1+R)}}$ is the sound speed and $R = \frac{3\rho_b}{4\rho_r} = 3a/4$ (when
assuming collisonless cold dark matter)
where $\rho_b$  and $\rho_r$ are the energy densities of baryons and radiation respectively \cite{Zhai_2019}.

Let us see what effect $\Tilde{\alpha}$ has on this equation. We notice that \eqref{constraintfried} gives us an expression for $H$:
\begin{equation}\label{eq:Hubblefact}
    H(a) = \frac{H_0}{\sqrt{2\Tilde{\alpha}}} \sqrt{\frac{2\alpha k}{a^2} - 1 + \sqrt{1+4\Tilde{\alpha} \left(\frac{\Omega_{m0}}{a^3}+\frac{\Omega_{r0}}{a^4}+\Omega_{\Lambda 0} + \frac{\tilde{\alpha} \tilde{C}^2}{a^4}\right)}}.
\end{equation}

As a first step, this then allows us to find an analytic solution to \eqref{eq:SoundHorizonLCDM} close to the Big Bang (in the limit $a \rightarrow 0$). Within GR we have the standard result
\begin{equation*}
    \begin{array}{lr}
        H \approx H_0\sqrt{\Omega_{r0}}a^{-2}\\
        R \rightarrow 0
    \end{array}
\Big\}
\implies r_s \approx \frac{c}{H_0}\int^a_0\frac{1}{\sqrt{3\Omega_{r0}}}\text{d}a'\approx \frac{1}{H_0} \frac{1}{\sqrt{3\Omega_{r0}}}a \rightarrow 0
\end{equation*}
whereas for a 4DEGB universe  
\begin{equation*}
    \begin{array}{lr}
        H \approx \mathcal{K} H_0 a^{-1}\\
        R \rightarrow 0 
    \end{array}
\Big\}
\implies r_s \approx \frac{1}{\sqrt{3}H_0\mathcal{K}} \int^a_0\frac{1}{a'}\text{d}a' \approx \frac{1}{\sqrt{3}H_0\mathcal{K}}\left[ \text{ln}a\right]^a_0    
\end{equation*}
where $\mathcal{K} = \sqrt[4]{\frac{ \Omega_{r0} + \Tilde{\alpha}\Tilde{C}^2}{\Tilde{\alpha}}}$ is a constant. This simplified version of Equation
\eqref{eq:SoundHorizonLCDM}
diverges close to the Big Bang for a 4DEGB universe. In general, in a 4DEGB universe $r_s$ diverges
logarithmically as $a\to 0$ for any $\alpha \neq 0$ if we use Equation \eqref{eq:SoundHorizonLCDM} to define the sound horizon.
This seems to suggest 4DEGB predicts an infinite comoving sound horizon as long as $\alpha \neq 0$. 

This effect is perhaps not as severe as it may seem at first glance. First, there will in general be a cutoff at small $a$ at the Planck time $t_{\textrm{Planck}} = 5.39 \times 10^{-44}\textrm{ s}$, since quantum gravitational effects will become important. More generally, recall that what is physically relevant is an effective sound horizon, defined as the distance traveled by a sound wave that has started propagating after the end of inflation. Waves that left their sources earlier than the end of inflation underwent incredibly
large Doppler shifts as the universe expanded, so it is reasonable to consider them to be completely unobservable in practice. If this is the case, the lower bound of the integral in \eqref{eq:SoundHorizonLCDM} is some scale factor at the end of inflation rather than at the beginning of the universe. This then gives us a new equation for our sound horizon: 
\begin{equation}\label{eq:soundhorizon}
    r_s(a) = \int^a_{a_{\text{min}}} \frac{c}{H(a')a'^2\sqrt{3(1+R)}}\text{d}a'
\end{equation}
where $a_{\text{min}}$ is close to the end of inflation, $a_{\text{min}} \approx 10^{-28}$.
We now encounter no divergences when calculating $r_s$. In GR+$\Lambda$CDM, the contribution to the integral from early times is negligible and the exact value of the lower bound isn't important. In 4DEGB, the contributions are large at early times and an exact value of this lower bound is needed to find the sound horizon.

We can solve this apparent issue by assuming that Big Bang Nucleosynthesis (BBN) has to occur at the same time (and be described by similar physics) in 4DEGB as in GR because of primordial abundance constraints.      
\begin{figure}[h!]
    \centering
    \includegraphics[width=1.0\textwidth, center]{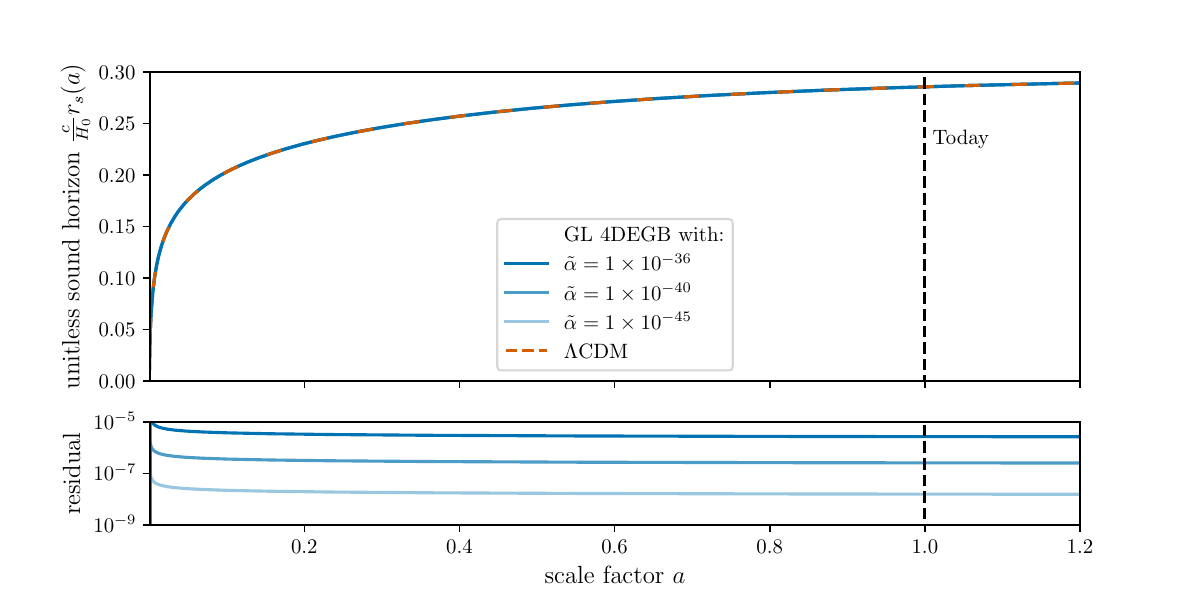}
    \vspace{-0.5cm}
    \caption{The dimensionless sound horizon as a function of redshift $z$ in GL 4DEGB compared to $\Lambda$CDM, with its residual. We assume fiducial cosmological parameters from \cite{2020} and $a_{\text{min}} \approx 10^{-28}$. We have set $C=0$ for convenience, but non-zero $C$ will not affect the sound horizon significantly. We display the sound horizon for values of $\alpha \lesssim 10^{17} \implies \Tilde{\alpha} \lesssim 10^{-36}$ \cite{Fernandes_2022}.}  
    \label{fig:soundhorizon}
\end{figure}
This requires $\alpha \lesssim 10^{17} \implies \Tilde{\alpha} \lesssim 10^{-36}$ \cite{Fernandes_2022}. We can then show that for this small value of $\Tilde{\alpha}$, if the end of inflation occurs around $a = a_{\text{min}} \approx 10^{-28}$, the exact lower bound of \eqref{eq:soundhorizon} isn't important. This small (but non-zero) $\alpha$ now causes a minimal modification to the sound horizon, which will be virtually the same in 4DEGB and  GR as illustrated in Figure \ref{fig:soundhorizon}.
Because the divergence of the integral in \eqref{eq:soundhorizon} is logarithmic, in general the lower cutoff does not meaningfully change the result. Even taking the cutoff $a_{\textrm{min}}$ to be  at the Planck time (before which any classical theory of gravity   is believed to be invalid), the difference between the $\Lambda$CDM and 4DEGB solutions would still be negligible for $\Tilde{\alpha} \lesssim 10^{-36}$.

In spite of this argument, having to impose a cutoff scale in the equation for the sound horizon is a theoretically unsatisfying characteristic of this theory. We have no firm knowledge of the correct $a_{\text{min}}$, and the fact that we have to pick one is a complication of the theory.

\subsection{The Horizon Problem}\label{horizonproblem}

Another interesting feature of 4DEGB in the early Universe is that in the limit of early time, the first equation in \eqref{eq:boxed_background_eq} reduces to:
\begin{equation}
    \alpha \MC{H}^4 =  \Omega_{r0} H_0^2 + \alpha C^2
\label{early_time_aH}
\end{equation}
and thus it follows that
\begin{equation}
    a(\eta) = a_{\text{min}}\,\text{exp}\left[H_0\sqrt[4]{\frac{ \Omega_{r0} + \Tilde{\alpha}\Tilde{C}^2}{\Tilde{\alpha}}}(\eta-\eta_{\text{min}})\right]=a_{\text{min}}\,\text{exp}\left[\MC{K}H_0(\eta-\eta_{\text{min}})\right]
\label{eq:earlyaH}
\end{equation}
where $\eta_{\text{min}}$ is the conformal time at which the cutoff scale factor $a_{\text{min}}$ occurs. 
In terms of comoving time $\eta$, \eqref{eq:earlyaH} means the universe isn't accelerating or decelerating ($\Ddot{a} = 0$), just expanding at a constant velocity. This implies that the comoving Hubble radius $(aH)^{-1}$ is a constant, as shown in Figure \ref{fig:particlehorizon}. 

\begin{figure}[h!]
    \centering
    \includegraphics[width=0.9\textwidth, center]{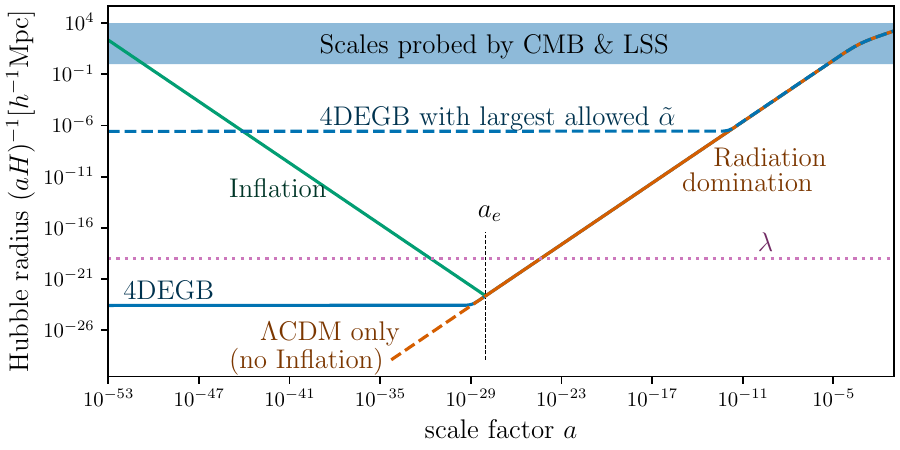}
    \vspace{-0.5cm}
    \caption{The comoving Hubble radius as a function of scale factor. Its logarithmic integral is the comoving particle horizon $\eta$. In general  $\MC{H}^{-1} = (a H)^{-1} \propto a^{n}$, 
    where for inflationary models $n=-1$, for radiation domination $n=1$ and in 4DEGB $n=0$. $a_e$ is the transition scale factor between inflation domination and radiation domination. Here $\lambda$ is an example of a primordial perturbation exiting the horizon during inflation and re-entering during radiation domination. We can see the same effect doesn't occur in 4DEGB. We assume fiducial cosmological parameters from \cite{2020} with $C=0$ and with $\Tilde{\alpha} = 10^{-112}$ and $\Tilde{\alpha} = 10^{-44}$ for the blue and blue-dashed lines respectively. The cutoff for the integration in \eqref{eq:PartHorizonLCDM} occurs at scale factors below the lowest limit of the x-axis for all investigated values of $\alpha$.}
\label{fig:particlehorizon}
\end{figure}

Let us now compare the comoving Hubble radius $(aH)^{-1}$ in 4DEGB to that in $\Lambda$CDM, and see the effect this has on the horizon problem. 
$\Lambda$CDM without inflation predicts that patches on the CMB are causally disconnected for $\theta \gtrsim 1$\textdegree, contrary to 
the observed uniformity of the CMB temperature map (which is 1 part in $10^5$). This is called the horizon problem. If the universe had an early epoch during which the comoving Hubble radius decreased or stayed constant, then there would be more  conformal time between the beginning of the universe and the CMB, so the past light cones of two points on the CMB would have had more time to be in causal contact. In the standard cosmological model, this is achieved by introducing an inflationary era at very early times. 

Alternatively, as mentioned above, 4DEGB also has this effect at early times (again, see Figure \ref{fig:particlehorizon}). As we see from Equation \eqref{eq:earlyaH}, 4DEGB is characterized by an early epoch where the Hubble radius $(aH)^{-1} = \dot{a}^{-1}$ is a constant. We then have a transition from an $\alpha$-dominated to a radiation-dominated universe. In this scenario, points in the CMB have overlapping past light cones and therefore originated from a causally connected region of space (see Figure \ref{fig:horizonlengthscales}).

\begin{figure}[h!]
    \centering
    \includegraphics[width=0.9\textwidth, center]{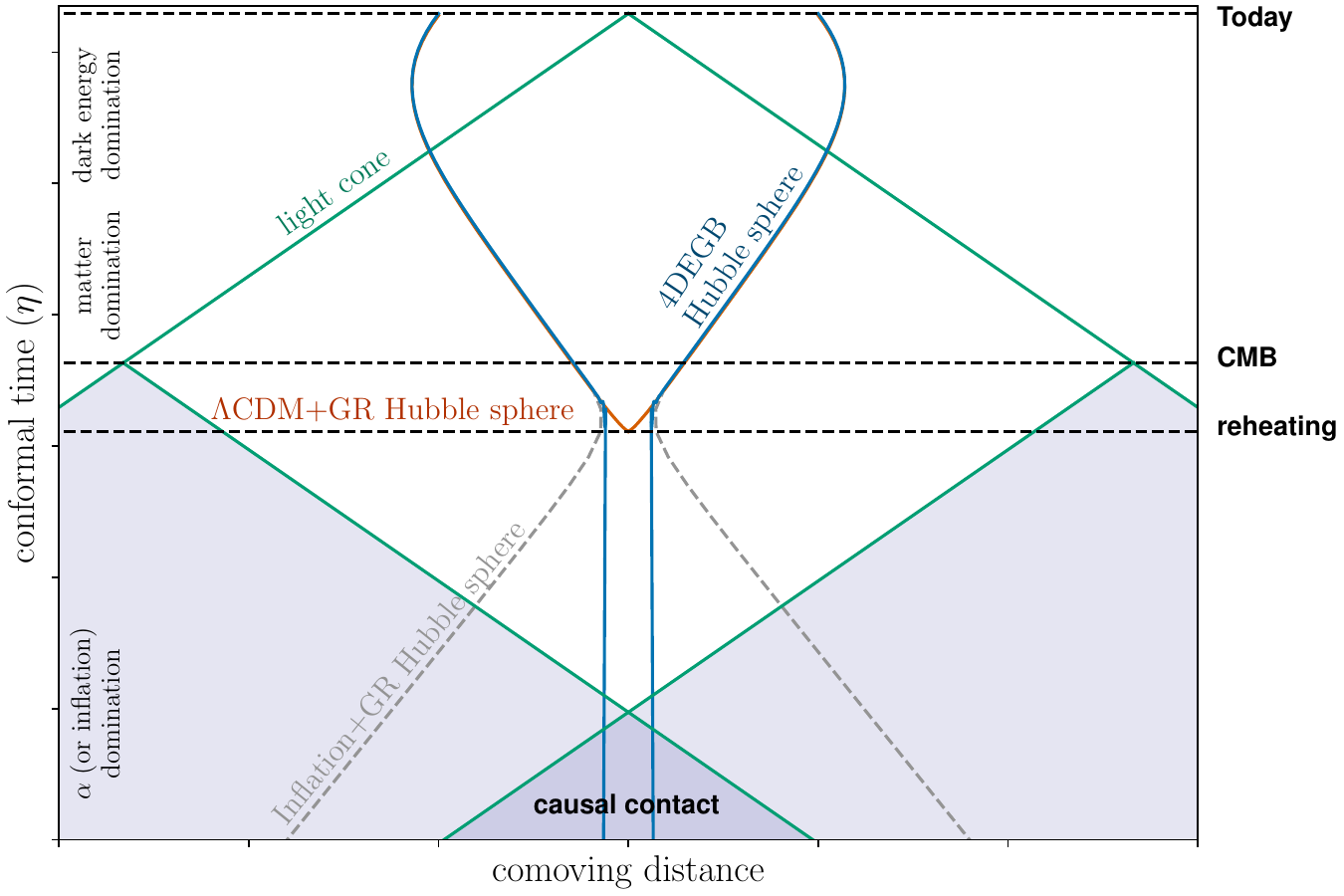}
    \vspace{-0.5cm}
    \caption{The 4DEGB solution compared to the inflationary solution to the horizon problem, visualized by looking at distances of interest as a function of conformal time ($\eta$). The plot assumes the exaggerated value of $\tilde{\alpha} = 10^{-7}$ to show the qualitative behaviour of the theory (and to aid the visual distinction between CMB and reheating space-like surfaces). We assume fiducial cosmological parameters from \cite{2020}. The $\Lambda$CDM+GR solution is plotted so that it matches 4DEGB at late conformal times. The reheating space-like surface is defined here as the conformal time at which the scale factor for the $\Lambda$CDM+GR solution goes to zero.}
    \label{fig:horizonlengthscales}
\end{figure}

Another way of thinking about this is by looking at the comoving particle horizon $\eta(z)$, defined as:
\begin{equation}\label{eq:PartHorizonLCDM}
    \eta(z) = \int^{z_{\text{max}}}_z \frac{c}{H(z')}\text{d}z' = \int^{a}_{a_{\text{min}}} \frac{c}{H(a')a'^2}\text{d}a'  = \int^{a}_{a_{\text{min}}} \frac{c}{H(a')a'
    }\text{dln}a'.
\end{equation}

This is the logarithmic integral of the inverse Hubble radius; it diverges logarithmically as $a_{\text{min}} \rightarrow 0$. A diverging comoving particle horizon (a diverging integral for \eqref{eq:PartHorizonLCDM}) means any two particles will have been in causal contact in some point in the past. However, as established above in Section \ref{Diverging horizons}, $a_{\rm min}$ must be finite and therefore Equation \ref{eq:PartHorizonLCDM} does not diverge. As a result, 4DEGB won't necessarily provide a solution to the horizon problem. Whether it does depends on how large the value of $\Tilde{\alpha}$ is. If we enforce $a_{\text{min}} =a(t_{\text{planck}})$, 4DEGB provides an independent solution to the horizon problem when $\Tilde{\alpha} \gtrsim 5 \times 10^{-11}$. Thus, while 4DEGB solves the horizon problem in principle, in practice this is a much larger $\Tilde{\alpha}$ than current constraints \eqref{eq:priors_alpha} allow.

4DEGB doesn't provide an early period of acceleration and $\Omega_k(z)$ is a constant in the early 4DEGB universe. Therefore, while 4DEGB can in principle be sufficient to solve the horizon problem, it can never solve the flatness problem. 4DEGB without inflation is also not sufficient to have primordial fluctuations exit and then re-enter the horizon 
as in $\Lambda$CDM+GR with inflation (again, see Figure \ref{fig:particlehorizon}). 
\section{Constraints from cosmological probes}\label{sec:LaterUniverse}

\subsection{A new formalism for cosmological parameter constraints}\label{sec:IntroCosmoConstr}

Although the cosmological impact of the parameter $\alpha$ alone is relatively weak other than in the very early Universe (see Section \ref{sec:VeryEarlyUniverse}), several studies have nevertheless placed constraints on $\alpha$ using late-time cosmological probes \citep{Wang_Mota, garciaaspeitia2021EGB}. While these do provide valuable independent constraints on $\alpha$, they are uninformative compared to bounds due to other effects. More importantly, we have seen in Section \ref{Diverging horizons} that to use observational information that relates to the sound horizon we must impose a strong prior from BBN of $|\alpha| \lesssim 10^{17}\text{m}^2$ \cite{Fernandes_2022}. This is the only meaningful way of using this information without assuming an inflationary model and doing a more detailed analysis of the sound horizon in 4DEGB. If we take this as a prior for a cosmological analysis using e.g. CMB and BAO data, the resulting constraints on $\alpha$ will then be completely prior-dominated.

Given the stringent constraints on $\alpha$ from Equation \eqref{eq:priors_alpha}, the GL formulation is very highly constrained. However, the full 4DEGB theory, with an additional free parameter $C$, has not been confronted with data (for one exception, see \citep{Feng_2021}, which discusses how one might constrain $C$ from theoretical considerations and existing observational bounds, but does not perform a full cosmological parameter inference analysis).
Motivated by the lack of conclusive constraints on the parameter $C$ in the literature, and by the stringent observational constraints on $\alpha$ as described above, we now introduce a version of the 4DEGB equations in the limit of small $\alpha$.

We develop the equations of 4DEGB in the limit where terms in $\alpha$, $\alpha A$, $\alpha A^2$ or $\alpha A^3$ are negligible compared to terms in $\alpha A^4 = \alpha C^2$. In this small-$\alpha$ limit and with $k=0$, Equation \eqref{eq:Hubblefact} takes the form:
\begin{equation}\label{eq:H_alphazero}
    H(z) = H_0\sqrt{\Omega_{m0} (1+z)^3 + (\Omega_{r0} + \Tilde{\alpha}_C)(1+z)^4 + \Omega_{\Lambda 0} }
\end{equation}
where we define a new 
dimensionless 
parameter
\begin{equation}
    \Tilde{\alpha}_C \equiv \Tilde{\alpha}\Tilde{C}^2
    \label{eq:alphaCdef}
\end{equation}
where $\tilde{C} \equiv CH_0^{-2}$, and $\alpha_C \equiv \alpha C^2$ such that $\tilde{\alpha}_C = \alpha_C H_0^{-2}$.

In the small-$\alpha$ limit, this will be the only parameter to enter both background and perturbation equations. We  assume $\tilde{\alpha}_C$ is finite and seek  to empirically constrain it. At times substantially past nucleosynthesis, the background is then equivalent to that of a $\Lambda$CDM universe with a ``dark radiation'' term that depends on $\tilde{\alpha}_C$.

We can now look at the set of equations \eqref{eq:boxed_pert_eq} in the same limit. These reduce to:

\begin{equation}{\label{eq:pert_eq_smallalpha}
    \begin{aligned}
        & k^2 \delta \varphi + 3\delta \varphi'' = 3(\Phi' + \Psi') \\
        & 12 \alpha_C \Phi - 12\alpha_C \delta \varphi' - 2a^2(k^2 \Psi + 3 \mathcal{H}\Psi' + 3 \mathcal{H}^2\Phi) = 8\pi Ga^4\delta \rho \\
        & \Psi' + \mathcal{H}\Phi = -4\pi G a^2 (1+\omega) \rho v \\
        & \Phi = \Psi + 8\pi G a^2 \Pi \\
        & 3 \Psi '' + 3\mathcal{H}(\Phi' + 2\Psi') + k^2(\Psi - \Phi) + (6 \mathcal{H}'-3\mathcal{H}^2)\Phi = 12\pi Ga^2 \delta \rho \\
        & \boldsymbol{\delta}^{\prime}=3 \mathcal{H} \boldsymbol{\delta}\left(w-\frac{\delta p}{\delta \rho}\right)+(1+w)\left(3 \Psi^{\prime}-\nabla^2 v\right) \\
        & 2 \nabla^4 \Pi+3 \nabla^2 \delta p+3(1+w) \bar{\rho}\left[(1-3 w) \nabla^2 v+\nabla^2 v^{\prime}+\nabla^2 \Phi\right]=0
    \end{aligned}
    }
\end{equation}
where $\delta \varphi = \frac{\delta \phi}{\sqrt{C}}$ is the same order of magnitude as $\Phi$ with respect to the constant $C$.  For a derivation of the equations above, we refer the reader to Appendix \ref{A1_perts_smallalpha}.

This limit of 4DEGB at the background level behaves like a $\Lambda$CDM universe with an additional contribution to the dark relativistic degrees of freedom. Interestingly, the parameter $\Tilde{\alpha}_C$ couples this background dark radiation term (which has a purely geometric origin) to the modified equations for structure growth. This offers the prospect of  differentiating 4DEGB from modifications to the standard model of particle physics that cause dark radiation, since the equations for growth will be different.

We can also look at equations (146c) and (146d) from \cite{Fernandes_2022} in the same limit. These reduce to:
\begin{equation}
    \begin{split}
        & a^2 \partial^2\Dot{F_i} + 16\pi G(1+\omega)a^4\Bar{\rho} v_i = 0 \\
    & a^2\Ddot{\gamma_{ij}\hspace{-6pt}}\hspace{6pt} + 2\mathcal{H}a^2 \Dot{\gamma_{ij}\hspace{-6pt}}\hspace{6pt} - a^2 \partial^2\gamma_{ij} = 8\pi G a^4 \Pi_{ij}.
    \end{split}
\end{equation}
These tell us that the tensor and vector perturbations reduce to those from $\Lambda$CDM in the small-$\alpha$ limit, even though the scalar perturbations are modified. Then any data that depends only on the vector and tensor perturbations (such as data from gravitational waves) won't enforce additional constraints on our parameter $\Tilde{\alpha}_C$.

In the following subsections we use cosmological probes of background equations to constrain our parameter $\tilde{\alpha}_C$. In Section \ref{perts_lateuniverse} we also introduce some qualitative effects of the modifications to the scalar perturbations on the power spectrum. 

\subsection{Background constraints on $\Tilde{\alpha}_C$ as a non-interacting dark radiation with a geometrical origin}\label{sec:alphaC_constraints_background}

Tests of  dark radiation have been carried out  extensively in the literature 
(see, e.g., \citep{Steigman_2007, Sch_neberg_2019,Riemer_S_rensen_2013, abbott2023dark, 2020}). We can take advantage of existing constraints and map them onto the question of constraining the $\tilde{\alpha}_C$ parameter of 4DEGB in the small-$\alpha$ limit. We first place background constraints by using datasets that don't require the assumption of a GR+$\Lambda$CDM structure growth history.

We will find it productive here and in the following section to examine how analyses that consider additional beyond-standard model neutrinos can equally constrain a cosmological model with standard model neutrinos and 4DEGB gravity, parameterized by $\tilde{\alpha}_C$ in the small-$\alpha$ limit. Flexibility in the number of such early-Universe relativistic degrees of freedom is often parameterized by:
\begin{equation}
    \Omega_{\nu\text{0}} = N_{\text{eff}}\frac{7}{8}\left( \frac{4}{11}\right)^{7/8} \Omega_{\gamma 0} 
\end{equation}
where $N_{\text{eff}}$ is the effective number of relativistic neutrino species (equal to 3.044 for standard model neutrinos only) and $\Omega_{\gamma 0}$ is the dimensionless photon density parameter. Constraints on $N_{\rm eff}$ from this $\Lambda$CDM+$N_{\text{eff}}$ model exist in the literature from a number of probes of the cosmological background dynamics (see, e.g., \cite{Steigman_2007, Sch_neberg_2019}), and we will take advantage of this here. 

At the level of the background only, we can treat $\tilde{\alpha}_C$ as if it were a contribution from an additional massless particle that neither interacts nor decays. Doing so, we can write down an equivalence between $\tilde{\alpha}_C$ in 4DEGB and $N_{\rm eff}$ in GR: 
\begin{equation}
    \tilde{\alpha}_C = (N_{\text{eff}}-3.044)\frac{7}{8}\left( \frac{4}{11}\right)^{7/8}\Omega_{\gamma 0}= \Delta N_{\text{eff}}\frac{7}{8}\left( \frac{4}{11}\right)^{7/8}\Omega_{\gamma 0} 
\end{equation}

It is important to emphasize that $\tilde{\alpha}_C$ has a geometrical origin --- we are not proposing to test a model with a varying number of neutrinos. We are testing a 4DEGB universe with standard model neutrinos (with $N_{\text{eff, SM}} = 3.044$). This means we will allow, for example, negative values of $\tilde{\alpha}_C$ which correspond to $N_{\text{eff}} < 3.044$, but the number of physical neutrino species $N_{\text{eff, SM}}$ will always be the standard 3.044. We emphasize also that this equivalence  should only be seen as a valid description of {\it background} cosmological observables; those that are sensitive to the growth of structure require consideration of modifications to the perturbation equations.

 With this in mind, we can initially consider constraints from $^4\mathrm{He}$ abundances (which give a conservative bound of $|N_{\text{eff}}| \leq 3.5$ \cite{Steigman_2007}). This maps to an upper bound on $\tilde{\alpha}_C$ of:
\begin{equation*}
    |\Tilde{\alpha}_C| \lesssim 10^{-5}.
\end{equation*}

\subsubsection{Constraints on $\Tilde{\alpha}_C$ from the CMB}\label{Late_universe}

Next, we will look at data from the Atacama Cosmology Telescope (ACT), specifically that taken with its second-generation polarization-sensitive receiver ACTPol \cite{thornton2016atacama}. This dataset provides information about the CMB  anisotropies, which in general depend on the perturbative behavior of the theory as well as the background behaviour. In practice, the effect of small modifications to the perturbation equations (which will change the Sachs-Wolfe and Integrated Sachs-Wolfe effects) is apparent only at angular multipoles $\ell \lesssim 200$ \cite{Intro_CMB_anisot}. The ACTPol analysis that we build on here takes a minimum multipole of 600 in the CMB temperature auto-spectrum, and 350 in the E-mode polarization auto-spectrum as well as the cross-spectrum between E-mode polarization and temperature, which means we can treat it effectively as a background-only probe. We use posterior chains of cosmological parameters obtained with the {\tt actpollite} likelihood for a $\Lambda$CDM+$N_{\text{eff}}$ model, which results in $N_{\text{eff}}= 2.42 \pm 0.43$ (using the maximum marginalized likelihood at a 68\% confidence level) \cite{Aiola_2020}. 

\begin{figure}[h!]
    \centering\includegraphics[width=0.7\textwidth, center]{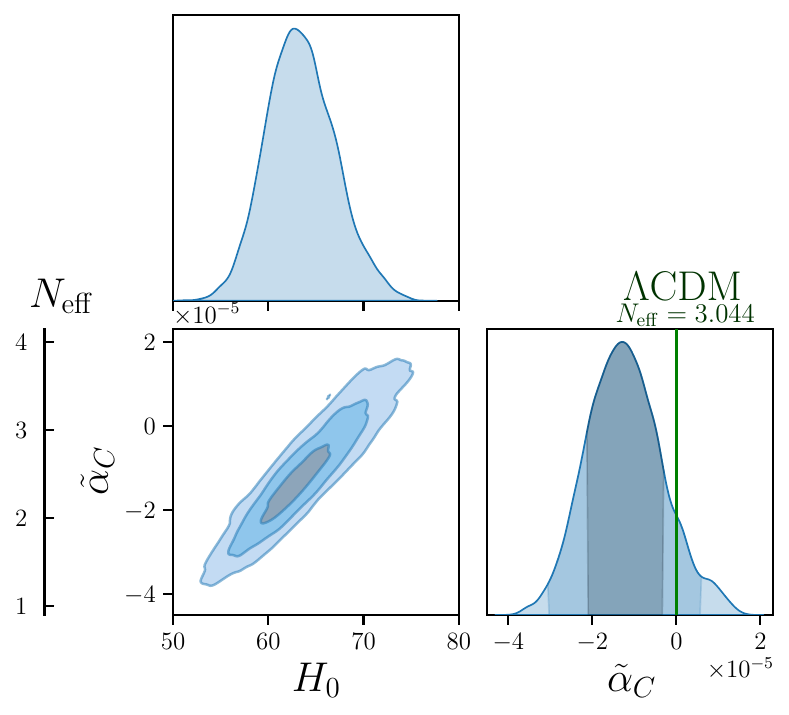}
    \vspace{-1cm}
    \caption{ ACTPol posterior results for a $\Lambda$CDM universe with varying $N_{\rm eff}$. We map this analysis onto our scenario of interest, with $N_{\rm eff}$ playing an equivalent role to $\tilde{\alpha}_C$ in 4DEGB for background-only probes. $\tilde{\alpha}_C$ is consistent with zero within $1.34\sigma$. We are displaying the $1\sigma$, $2\sigma$ and $3\sigma$ confidence regions in the contour plot, and the $1\sigma$ and $2\sigma$ regions in the 1D posterior probability distribution for $\tilde{\alpha}_C$.
    }
    \label{fig:cornerplot}
\end{figure}

This translates to a constraint for 4DEGB (with standard neutrinos) of:
\begin{equation}
    \Tilde{\alpha}_C = (-3.5 \pm 2.4)\times 10^{-6}h^2 = (-9 \pm 6) \times 10^{-6}
\end{equation}
at 68\% confidence level, which is consistent with zero at the 1.34$\sigma$ level. $h$ is the dimensionless Hubble constant defined by $H_0 = 100 h \textrm{ km}\hspace{0.2em}\textrm{s}^{-1}\textrm{Mpc}^{-1}$ and we have used the maximum posterior value for $h$ from the ACTPol analysis in question in our calculations.

As we can see in Figure \ref{fig:cornerplot}, the analysis of \cite{Aiola_2020} prefers a lower $N_{\text{eff}}$ than the standard model (or in 4DEGB a negative $\Tilde{\alpha}_C$). An interesting feature of 4DEGB is that it provides a modified gravity framework that mimics a continuous variation of $N_{\text{eff}}$ from its $\Lambda$CDM value of 3.044, including values $N_{\text{eff}} < 3.044$. Few models are well-motivated theoretically for $N_{\text{eff}} < 3.044$ \cite{2020}, so 4DEGB provides an interesting new precedent, which should be borne in mind should future constraints also favour a negative value for $\Delta N_{\rm eff}$. While we cannot draw conclusions from all-sky CMB Planck data without accounting for a modification to the scalar perturbation equations, we note here for completeness that this preference for a lower value of $N_{\text{eff}}$ does not seem to persist in Planck data \cite{2020} for a $\Lambda \text{CDM+}\text{GR}$ model.

\subsection{Effect of modified perturbations in 4DEGB on the matter power spectrum}\label{perts_lateuniverse}

Having considered the background-only effect of $\tilde{\alpha}_C$, we now examine the growth of perturbations in this small $\alpha$ limit of 4DEGB. We incorporate the 4DEGB background and perturbation equations within a bespoke cosmological Boltzmann code to study the qualitative behaviour of small-$\alpha$ 4DEGB in a perturbed universe. We use a modified version of equations \eqref{eq:H_alphazero} and \eqref{eq:pert_eq_smallalpha} (see Appendix \ref{A_boltzsolve}) to recover the matter power spectrum $P(k)$. The Boltzmann code we use is constructed following the methodology of \citep{callin2006calculate}. In taking this approach, we prioritize ease of implementation rather than the computational efficiency which would be gained by modifying the highly optimised public packages {\tt CAMB} \citep{Lewis:1999bs} or {\tt CLASS} \citep{Diego_Blas_2011}. We validate our Boltzmann code against these packages in the GR case and find excellent agreement.

\begin{figure}[h!]
    \vspace{-0.5cm}
    \centering
    \includegraphics[width=0.82\textwidth, center]{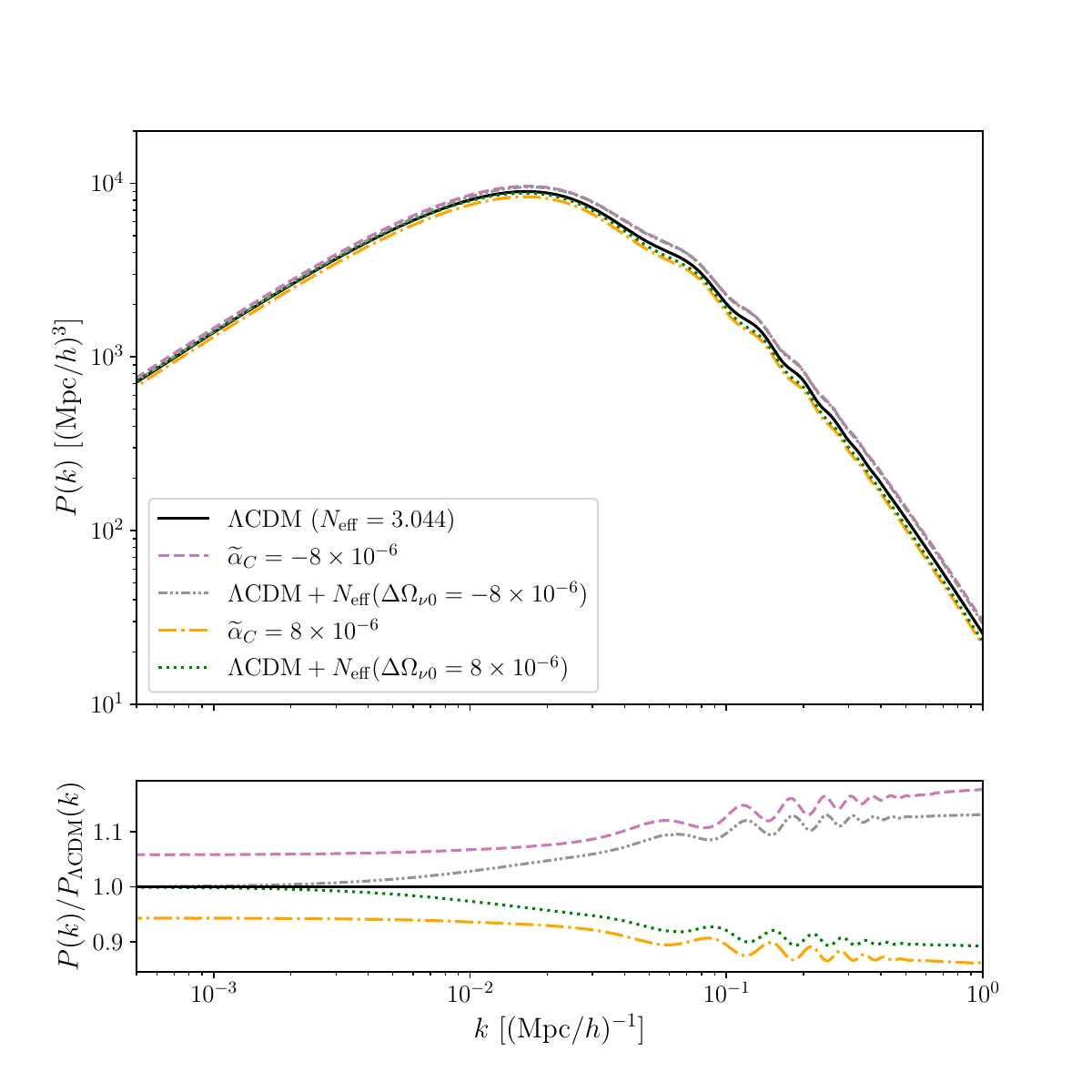}
    \vspace{-1.0cm}
    \caption{Present-day ($z=0$) linear matter power spectrum for small-$\alpha$ 4DEGB. The middle (solid black) curve is the power spectrum for a $\Lambda$CDM model. The values of $\Tilde{\alpha}_C$ displayed here are chosen to be of the same order of magnitude as background constraints of Section \ref{sec:alphaC_constraints_background}. The dotted (green) and dash-double dotted (gray) curves show the $\Lambda$CDM$+N_{\textrm{eff}}$ model with extra neutrino densities equivalent to the plotted values of $\Tilde{\alpha}_C$ ($\Delta \Omega_{\nu 0} = \Tilde{\alpha}_C = 8\times 10^{-6}$ and $-8\times 10^{-6}$ respectively), which correspond to $N_{\textrm{eff}} = 5.46$ and $0.62$ respectively. When varying $N_{\rm eff}$, we fix $\{Y_p, \Omega_M^0, \Omega_B^0, \Omega_R^0, h, A_s , n_s\}$ to be the fiducial cosmological parameters from \cite{2020}. We approximate all neutrinos as massless.}
    \label{fig:powerspec}
\end{figure}

The resulting power spectra can be seen in Figure \ref{fig:powerspec}. We see that the difference in $P(k)$ between 4DEGB and $\Lambda$CDM with the equivalent $N_{\textrm{eff}}$ value is non-negligible for $\Tilde{\alpha}_C \approx \pm 8 \times 10^{-6}$. The difference in the paired spectra is larger at large scales (low $k$); to understand this, we turn to the second equation in \eqref{eq:pert_eq_smallalpha}, which suggests that the $\tilde{\alpha}_C$ term is more dominant when $k$ is smaller. Beyond this large-scale effect, there is also a scale-dependent deviation between the two theories as we move to smaller scales (larger $k$). Combined, these effects suggest that there is potential for the degeneracy between this sector of 4DEGB and $\Lambda$CDM$+N_{\textrm{eff}}$ to be broken by observational probes sensitive to changes in the power spectrum (such as weak gravitational lensing, galaxy clustering, CMB spectra at lower multipoles which probe perturbation effects, or CMB lensing). Although the fractional deviations between the two models in Figure \ref{fig:powerspec} are small, upcoming surveys  such as {\it Euclid} \citep{Euclid2022} and the Rubin Observatory Legacy Survey of Space and Time (LSST, \citep{Ivezic2019}) may be capable of observationally distinguishing these percent-level effects. Such an analysis is beyond the scope of this paper but is of great interest for future work.

\section{Conclusion}
\label{sec:conc}

4DEGB   gravity has  broad theoretical interest  as a   competitor  to  general relativity, primarily because it is a 4-dimensional  higher-curvature theory of gravity that enjoys all the basic features of higher-dimensional Lovelock gravity.
It has also  provoked substantial observational interest as a theory with few extra parameters compared to $\Lambda$CDM$+$GR, which has resulted in the Glavin-Lin version being extensively investigated, and its single parameter $\alpha$ being well-constrained. 
Here we have for the first time studied the implications of   
scalar-tensor 4DEGB, providing the first observational constraints on a geometric term parametrized by a constant $C$.

Considering a background spacetime that is homogeneous and isotropic and a perturbed spacetime that only depends on scalar functions, we find that the stress-energy perturbations are that of a perfect fluid with anisotropic stress. 
This in turn dictates that there are 7 unknown functions 
correcting the cosmic background.
We have provided  the first derivation of the complete set of equations necessary for computing these 7 unknown functions  for both flat and curved spatial hyper surfaces. The conservation of stress-energy provides two equations of motion (that do not differ from GR), while the gravitational and  scalar field equations provide the other five. The complete set forms a nonlinear coupled PDE system that does not have a closed form solution. These equations become the first order perturbation equations of general relativity in the appropriate limit.

 We have explored some consequences of $\alpha$ in the very early universe, where the small, allowable values of this parameter can still have important cosmological effects. We find that the sound horizon in 4DEGB depends on the behaviour of the universe at early times, meaning that the value of the parameter $\alpha$ and the exact redshift at which sound propagation begins to matter both affect the measured solution for the sound horizon. We show that this issue can be alleviated somewhat, though this requires imposing a minimum scale factor after which 4DEGB becomes relevant. Furthermore, we have shown that 4DEGB can resolve the horizon problem: the particle horizon diverges at early times, meaning that all particles will have been in causal contact at some very early time. This solution fails for very low values of $\alpha$ if we enforce a cutoff to this classical theory at the Planck time. This is because the aforementioned divergence is logarithmic and therefore is very slow to diverge even for a very small minimum time. This theory also fails to act as a replacement for an inflation-type model in the early Universe because it does not resolve the flatness problem.

Unlike the  Glavin-Lin formulation, full 4DEGB contains a dark radiation-like term at the background level, whose associated parameter $C$  has remained largely unconstrained in the literature before now. Introducing a new parameter $\tilde{\alpha}_C = \frac{\alpha_C}{H_0^2}= \frac{\alpha C^2}{H_0^2}$ we have shown that, under the assumption that   $\alpha$ is small (justified by prior constraints on this parameter \cite{Fernandes_2022}), the equations describing the theory at times substantially past inflation depend only on $\Tilde{\alpha}_C$. The late-time behaviour of the theory can then be constrained as a combination of background ``dark radiation'' effects and a modification to structure growth. We have shown that the 4DEGB parameter is constrained using CMB data from ACTPol to 
\begin{equation}
    \Tilde{\alpha}_C = \Tilde{\alpha}\Tilde{C}^2 = (-9 \pm 6) \times 10^{-6} \implies \alpha_C = (-10 \pm 7) \times 10^{-58} h^2 \text{ m}^{-2}\approx (-4 \pm 3) \times 10^{-58} \text{ m}^{-2}
\end{equation}
where we have used the conversion factor $H_0^2 = (1.17\times10^{-52})h^2 \text{ m}^{-2}$. This implies $\alpha_C$ is consistent with zero within $1.34\sigma$. Interestingly, ACTPol data seems to prefer a lower value of $N_{\textrm{eff}}$ than that suggested from the standard model  (or in 4DEGB a negative $\Tilde{\alpha}_C$). 4DEGB enjoys the interesting feature of being a geometrical alternative to continuously varying $N_{\text{eff}}$, including values of $N_{\text{eff}} < 3.044$ that are much less theoretically motivated in other settings. This may prove interesting in the future, should further cosmological observables favour such values of $N_{\rm eff}$ in a $\Lambda$CDM+$N_{\text{eff}}$ model.

While we completed a qualitative analysis of the power spectrum in this paper, modifying an efficient public Boltzmann solver to compute this quantity would enable further constraints on $\alpha_C$ using observables sensitive to cosmic structure growth. This would enable us to break the degeneracy between this branch of 4DEGB and GR+$\Lambda$CDM with a varying $N_{\rm eff}$. Figure \ref{fig:powerspec} shows that the power spectrum (and therefore $\sigma_8$ and $S_8$) will be modified in 4DEGB compared to $\Lambda$CDM (or $\Lambda$CDM+$N_{\textrm{eff}}$). It would therefore be interesting to investigate in future work what impact this sector of 4DEGB might have on the $S_8$ tension. 

Finally, we note that there are other potential avenues of exploration of the cosmological consequences of 4DEGB in the very early universe. We have not here discussed the implications of a 4DEGB universe in terms of the behaviour and evolution of super-horizon perturbations, or the implications regarding early-universe singularities or lack thereof in these theories. A universe dominated by 4DEGB at early times might have an effect on the scale dependence or non-gaussianity of the primordial power spectrum which would be interesting to investigate. As yet no such calculation has been attempted. The work presented here on the sound horizon and its divergent behaviour as $a \rightarrow 0$ suggests that these may be fruitful avenues for further work.

\acknowledgments

This work was supported in part by the Natural Sciences and Engineering Research Council of Canada and in part by the Philip Robinson Cosmology PhD studentship.
We thank Ian Moss and Ghazal Geshnizjani for helpful discussions, and Pedro G. Ferreira for guidance in writing the original GR version of the Boltzmann code modified for this work. We also thank Chris Harrison for kindly allowing the
use of his research computing server and Erminia Calabrese for providing helpful assistance with accessing ACTPol data products.
The Python libraries SciPy \cite{2020SciPy-NMeth}, NumPy \cite{harris2020array},
PyCCL\footnote{https://github.com/LSSTDESC/CCL} \cite{Chisari_2019} and corner.py \cite{corner} were significant
in enabling this work to be done. 

\noindent
Author contributions: CMAZ performed the analysis of Sections \ref{sec:VeryEarlyUniverse} and \ref{sec:LaterUniverse} (including writing the corresponding code) and wrote a significant portion of the text of this paper. BRH conducted the theoretical calculations of Section \ref{Sec:Theory} and wrote a significant portion of the text of this paper. CDL provided technical and conceptual guidance and expertise most pertinent to Sections \ref{sec:VeryEarlyUniverse} and \ref{sec:LaterUniverse}, checked some calculations, wrote the original GR version of the Boltzmann code used in Section \ref{sec:LaterUniverse}, wrote some text for this paper, and revised the paper text. RBM provided technical and conceptual guidance and expertise most pertinent to Section \ref{Sec:Theory}, checked some calculations, wrote some text for this paper, and revised the paper text.

For the purpose of open access, the author has applied a Creative Commons Attribution (CC BY) licence to any Author Accepted Manuscript version arising from this submission.

\appendix

\section{Equations of Motion for $k \ne 0$ spatial sections}\label{Appendix:equation_of_motion_curvature}
Here we will present the gravitational and scalar field equations for non-flat ($k\ne 0 $) spatial curvature where we compute in conformal time with the line element outlined by \eqref{FRWCON}. While some of the presentations will be redundant,  we nonetheless will proceed with clarity in our calculations. The background calculations are carried out in Section \ref{backgroundsection} so we will not repeat them here, and instead  just note that the solution to \eqref{scalarfieldequation} is given as 
\begin{equation}
    \bar{\phi}^{\prime}=\mathcal{H}+A \Rightarrow \bar{\phi}=\ln (a)+A \eta+B
\end{equation}
which used in conjunction with \eqref{fieldequations} yields the finalized gravitational field equations
\begin{equation} \label{zeroset}
\begin{aligned}
& \frac{\left(\mathcal{H}^2+k\right)^2 \alpha}{a^2}+\mathcal{H}^2+k=\frac{8 \pi G a^2 \bar{\rho}}{3}+\frac{\alpha C^2}{a^2} \\ 
& \frac{\left(\mathcal{H}^2+k\right)\left(4 \mathcal{H}^{\prime}-\mathcal{H}^2-k\right) \alpha}{a^2}+\mathcal{H}^2+2 \mathcal{H}^{\prime}+k=-8 \pi G a^2 \bar{p}-\frac{\alpha C^2}{a^2} 
\end{aligned}
\end{equation}
where we have made the substitution $A^2=-k+C$. The conservation of stress-energy $\nabla_\mu T^{\mu \nu}=0$ yields only one equation for the background, that being
\begin{equation}
    \bar{\rho}^{\prime}=-3 \MC{H}(\bar{p}+\bar{\rho})
\end{equation}

For the first order equations we will present first with the arbitrary constant$A$ and then make the substitution $A^2=-k+C$,  placing a box around the finalized equation. Beginning with the scalar field equation,which is found in Section \ref{Theory:PerturbedSpacetime}
\begin{equation}
\alpha\left(A^2+k\right)\left(\nabla^2(\delta \phi-\Phi)+3 A\left(\Psi^{\prime}+\Phi^{\prime}\right)-3\left(\Psi^{\prime \prime}+\delta \phi^{\prime \prime}\right)\right)=0
\end{equation}
\begin{equation}
  \boxed{  \alpha C \left(\nabla^2(\delta \phi-\Phi)+3 \sqrt{-k+C}\left(\Psi^{\prime}+\Phi^{\prime}\right)-3\left(\Psi^{\prime \prime}+\delta \phi^{\prime \prime}\right)\right)=0.}
\end{equation}
For the gravitational field equations, we will consider the $\eta \eta$ component first. The stress energy tensor remains the same $\delta T_{\eta \eta}=a^2 \delta \rho+2 a^2 \Phi \bar{\rho}$. Computing $\delta \MC{E}_{\eta \eta}= 8 \pi G( a^2 \delta \rho+2 a^2 \Phi \bar{\rho})$ and using the first equation of \eqref{zeroset} we arrive at the simplified expression
\begin{equation}
    \begin{aligned}
    & 4 \alpha A^2 \left[ 3 A^2 \Phi - 3 A( \delta \phi^{\PR} + \Psi^{\PR}) - \lap(\delta \phi + \Psi) - 3 k (\Psi - \Phi) \right] - 4 \alpha k \left[ 3 A (\delta \phi^{\PR} + \Psi^{\PR}) + \lap \delta \phi \right] + 2 \MC{A} \lap \Psi \\
    & - 6 \MC{H}( \MC{A} + 2\alpha k) \Psi^{\PR} - 12 \alpha \MC{H}^2 \left((k+\MC{H}^2)\Phi - k \Psi \right) + 6 a^2 ( k \Psi - \MC{H}^2 \Phi) = 8 \pi G a^4 \delta \rho.
    \end{aligned}
\end{equation}
\begin{equation*}
\mathclap{
    \boxed{\begin{aligned}
        & 2 ( \MC{A}+2 k \alpha) ( \lap \Psi - 3 \MC{H} \Psi^{\PR} + 3 (k \Psi - \Phi \MC{H}^2)) + 4 \alpha C( 3 \Phi C  - \lap(\Psi + \delta \phi) - 3 \sqrt{-k+C}(\delta  \phi^{\PR}+\Psi^{\PR})) = 8 \pi a^4 G \delta \rho.
    \end{aligned}}}
\end{equation*}
Any one of the time-spatial equations $\delta \MC{E}_{0i} = 8 \pi G \delta T_{0i} $ has no corresponding zeroth order equation so calculating this directly yields
\begin{equation*}
\mathclap{
    4 \alpha A^2 \left[ A (\Phi_{,i} - \delta \phi_{,i}) - ( \delta \phi^{\PR}_{,i} + \psi^{\PR}_{,i})\right] + 4 \alpha k A (\Phi_{,i}-\delta \phi_{,i}) + 2 \MC{H}( \MC{A} + 2 \alpha k) \Phi_{,i} + 2 \MC{A} \Psi^{\PR}_{,i} - 4 \alpha k \delta \phi^{\PR}_{,i} = - 8 \pi G a^4(1+w) \bar{\rho} v_{,i}}
\end{equation*}
\begin{equation}
    \boxed{2 \MC{H}(\MC{A}+2 \alpha k)(\Phi + \Psi^{\PR}) + 4 \alpha C ( \sqrt{-k+C}(\Phi - \delta \phi) - \delta \phi^{\PR} - \Psi^{\PR}) = - 8 \pi G (1+w) a^4 \bar{\rho} v.}
\end{equation}
For a mixed spatial-spatial component $\delta \MC{E}_{ij} = 8 \pi G \delta T_{ij}$ for $i \ne j$ we have
\begin{equation}
    \nabla_{i}\partial_{j}( 2 \alpha A^2 ( \Phi + \Psi) + 4 \alpha \MC{H}^{\PR} \Psi - \MC{A} \Phi - \MC{B} \Psi) = 8 \pi G a^4 \nabla_{i}(\partial_j \Pi)
\end{equation}
\begin{equation}
     \boxed{2 \alpha (-k+C) ( \Phi + \Psi) + 4 \alpha \MC{H}^{\PR} \Psi - \MC{A} \Phi - \MC{B} \Psi = 8 \pi G a^4 \Pi}.
\end{equation}
For the diagonal spatial components we employ the same procedure as with the flat case. We consider all the components individually and compute the trace, $\delta \MC{E}_{rr} +\delta \MC{E}_{\theta \theta }+\delta \MC{E}_{\phi \phi} = 8 \pi G ( \delta T_{rr} + \delta T_{\theta \theta } + \delta T_{\phi \phi}) $; using the 2nd equation of \eqref{zeroset}, this gives
\begin{equation}
    \begin{aligned}
        & 4 \alpha A^2 \left( 3 A^2 \Phi + 3 A ( \Phi^{\PR} - \delta \phi^{\PR}) - \lap( \Phi + \Psi) - 3 (\Psi^{\PR \PR }+\delta \phi^{\PR \PR }) + 3 k (\Phi - \Psi)\right) + 12 k \alpha A (\Phi^{\PR}- \delta \phi^{\PR}) \\
        & + 4 \MC{H}^{\PR} \left( 6 \MC{H} \alpha \Psi^{\PR} + 3 ( 4 \MC{H}^2 + a^2 + 2 k \alpha )\Phi - 6 k \alpha \Psi - 2 \lap \Psi \alpha \right) + 2 \MC{A} \lap \Phi + 6 \MC{H}(\MC{A} + 2 k \alpha)\Phi^{\PR} + 2 \MC{B} \lap \Psi \\
        & -12 \alpha k \delta \phi^{\PR \PR} + 6 \MC{A} \Psi^{\PR \PR} + 12 \MC{H} a^2 \Psi^{\PR} - 12 \MC{H}^2( \Phi \MC{H}^2 + k(\Phi - \Psi)) + 6 a^2(\MC{H}^2 \Phi - k \Psi)= 24 \pi G a^4 \delta p
    \end{aligned}
\end{equation}
\begin{equation*}
\mathclap{
    \boxed{\begin{aligned}
       & (\MC{B} + 2k \alpha) \lap \Psi + (\MC{A} + 2k \alpha) (\lap \Phi +3 \Psi^{\PR \PR}) + 2 \alpha C \left( -\lap \Phi - 3(\delta \phi^{\PR \PR}+\Psi^{\PR \PR}) - 3 k( \Phi + \Psi) - 3 \sqrt{-k+C}\delta \phi^{\PR} \right) \\
       & 2 \MC{H}^{\PR} \left( - 2 \alpha  \lap \Psi + 12 \alpha \MC{H} \Phi + 3 a^2 \Phi + 6 \alpha k \Phi - 6 \alpha k \Psi + 6 \alpha \MC{H} \Psi^{\PR} \right) + 6 \alpha C^2 \Phi + 6 \MC{H} a^2 \Psi^{\PR} + 3 a^2(\MC{H}^2-k\Psi) \\
       & + 3 \left( 2 \alpha \MC{H}^3 + 2 \alpha (-k+C)^{3/2} + \MC{H}a^2 + 2 \MC{H} \alpha k + 2 k \alpha \sqrt{-k+C} \right) \Phi^{\PR} - 6 \alpha (\MC{H}^2+k)(\MC{H}^2 \Phi - \Psi k) = 12 \pi G a^4 \delta p.
    \end{aligned}}}
\end{equation*}
We now need to consider the conservation of stress energy, $\nabla_{\mu}T^{\mu \nu}=0$. Considering the time component first $\nabla_{\mu}T^{\mu \eta} =0$, we obtain the same zeroth and first order equations as in the flat case; hence
\begin{equation}
    \boxed{\delta \rho^{\prime}=(1+w)\left(3 \Psi^{\prime}-\nabla^2 v\right) \bar{\rho}-3 \mathcal{H}(\delta \rho+\delta p)}\; .
\end{equation}
For the spatial components there are some slight modifications. For any spatial coordinate $i$, the equation $\nabla_{\mu}T^{\mu i}=0$ becomes 
\begin{equation}
    \partial_{i}\left( \lap \Pi + (3 w+3) \bar{\rho} v^{\PR} -3(1+w)(3 w-1) \MC{H} \bar{\rho}v + 3(1+w) \bar{\rho} \Phi + 3 \delta p + 6 k \Pi \right)=0.
\end{equation}
Taking the spatial divergence of this yields 
\begin{equation}
    \nabla^{i} \left(\partial_{i}\left( \lap \Pi + (3 w+3) \bar{\rho} v^{\PR} -3(1+w)(3 w-1) \MC{H} \bar{\rho}v + 3(1+w) \bar{\rho} \Phi + 3 \delta p + 6 k \Pi \right)\right)=0
\end{equation}
\begin{equation}
    \lap\left( \lap \Pi + (3 w+3) \bar{\rho} v^{\PR} -3(1+w)(3 w-1) \MC{H} \bar{\rho}v + 3(1+w) \bar{\rho} \Phi + 3 \delta p + 6 k \Pi \right)=0
\end{equation}
{\small
\begin{equation}
    \boxed{\nabla^4 \Pi + 3(1+w)\bar{\rho} \lap v^{\PR}-3(1+w)(3 w-1) \mathcal{H} \bar{\rho} \lap v+3(1+w) \bar{\rho} \lap \Phi + 3 \lap \delta p + 6k \lap \Pi =0}
\end{equation}}

\section{Scalar Perturbations in the small-$\alpha$ limit}\label{A1_perts_smallalpha}

We start with equations \eqref{eq:boxed_pert_eq}. We derive an approximation for these scalar perturbations in the following limits:

\begin{enumerate}
    \item $\alpha$ is small, meaning $\alpha/H_0^2 << 1$. This is a reasonable approximation given \citep{Clifton_2020} gives us priors 
    $-10^{-83} \lesssim \Tilde{\alpha} \lesssim 10^{-43}$. We can then write $\mathcal{A} = \mathcal{C} = \mathcal{D} = a^2$ and $\mathcal{B} = -a^2$.
    \item $C$ is large. Specifically, we assume large enough that $\alpha_C = \alpha C^2$ is finite and affects late universe dynamics, but small enough that $\alpha C^{1/2}$, $\alpha C$ and $\alpha C^{3/2}$ terms can be neglected.
    \item $\alpha$ is small enough that the background equation reduces to $H^2 = \frac{8\pi G\rho}{3} + \frac{\alpha_C}{a^4}$.
\end{enumerate}

Using these approximations we can rewrite the first equation in \eqref{eq:boxed_pert_eq} as:
\begin{equation}\label{eq:dphi}
    k^2 \delta \phi + 3\delta \phi'' = 3\sqrt{C}(\Phi' + \Psi')
\end{equation}
We can do this because we know $\nabla^2\Phi << \sqrt{C}\Phi'$ and $\Psi'' << \sqrt{C}\Psi'$. This suggests that $\delta \varphi = \frac{\delta \phi}{\sqrt{C}}$ is the same order of magnitude as the metric perturbations $\Phi$ and $\Psi$, so that we can re-write our equation as:
\begin{equation}\label{eq:dphibar}
    k^2 \delta \varphi + 3\delta \varphi'' = 3(\Phi' + \Psi')
\end{equation}
Then using the same approximations we can rewrite the second equation in \eqref{eq:boxed_pert_eq} as:
\begin{equation}\label{eq:poissoneq}
    12 \alpha_C \Phi - 12\alpha_C \delta \varphi' - 2a^2(k^2 \Psi + 3 \mathcal{H}\Psi' + 3 \mathcal{H}^2\Phi) = 8\pi Ga^4\delta \rho
\end{equation}

The remaining equations reduce to their GR equivalent:
\begin{equation}{\label{eq:pert_eq_GR}
    \begin{aligned}
        & \Psi' + \mathcal{H}\Phi = -4\pi G a^2 (1+\omega) \rho v \\
        & \Phi = \Psi + 8\pi G a^2 \Pi \\
        & 3 \Psi '' + 3\mathcal{H}(\Phi' + 2\Psi') + k^2(\Psi - \Phi) + (6 \mathcal{H}'-3\mathcal{H}^2)\Phi = 12\pi Ga^2 \delta \rho \\
        & \boldsymbol{\delta}^{\prime}=3 \mathcal{H} \boldsymbol{\delta}\left(w-\frac{\delta p}{\delta \rho}\right)+(1+w)\left(3 \Psi^{\prime}-\nabla^2 v\right) \\
        & 2 \nabla^4 \Pi+3 \nabla^2 \delta p+3(1+w) \bar{\rho}\left[(1-3 w) \nabla^2 v+\nabla^2 v^{\prime}+\nabla^2 \Phi\right]=0
    \end{aligned}
    }
\end{equation}
\section{Coupled Differential Equations for the Boltzmann Solver}\label{A_boltzsolve}

The following derivations follow the procedures outlined in \cite{callin2006calculate,Dodelson}.

The Boltzmann equations for photons, photon polarization, neutrinos, dark matter and baryons are independent of the cosmological model (they describe the equations in any given expanding spacetime with perturbations). They will therefore be the same in $\Lambda$CDM and 4DEGB. These can be written as (see equation (22) in \cite{callin2006calculate}):
\begin{equation}\label{eq:A_boltzmann}
    \begin{split}
        &\Theta' + ik\mu\Theta = \Psi' -ik\Phi - \tau'\left[\Theta_0 - \Theta + ik\mu v_b - \frac{1}{2}\mathcal{P}_2 \Pi\right]\\
        &\Theta_P' +ik\mu \Theta_P = -\tau'\left[\Theta_P +\frac{1}{2}(1-\mathcal{P}_2)\Pi\right] \\
        & \delta'-k^2v = 3\Psi' \\
        &v' +\mathcal{H}v =-\Phi \\
        &\delta_b' -k^2v_b = 3\Psi' \\
        & v_b' +\mathcal{H}v_b = -\Phi +\frac{\tau' R}{k}(v_b + 3\Theta_1) \\
        &\mathcal{N}' +ik\mu \mathcal{N} = \Psi' -ik\mu \Phi
    \end{split}
\end{equation}
where all quantities are defined as in \cite{callin2006calculate}, except for $\Phi_{\textrm{\cite{callin2006calculate}}} = -\Psi$, $\Psi_{\textrm{\cite{callin2006calculate}}} = \Phi$ and $v_{\textrm{\cite{callin2006calculate}}} = kv$. Care should also be taken when comparing derivatives (the prime symbol in \cite{callin2006calculate} indicates a derivative with respect to $x=\textrm{ln}a$). To explicitly solve the coupled ODEs we need to use the multipole expansion of the first two equations in \eqref{eq:A_boltzmann}.

Next, by looking at \eqref{eq:poissoneq} and the second equation in \eqref{eq:pert_eq_GR} (when including photon and neutrino perturbations) we find that the relevant Einstein equations take the form:
\begin{equation}\label{eq:A_einstein}
    \begin{split}
        & k^2\Psi + 3\mathcal{H}(\Psi' + \mathcal{H}\Phi) = \frac{6\alpha_C(\Phi - \delta \varphi')}{a^2}  + 4\pi Ga^2 \left[ \rho\delta + \rho_b \delta_b+4\rho_{\gamma}\Theta_0+4\rho_{\nu}\mathcal{N}_0\right]\\
        &\Phi = \Psi -\frac{12H_0^2}{k^2a^2}\left[\Omega_{\gamma 0}\Theta_2 + \Omega_{\nu 0}\mathcal{N}_2\right]
    \end{split}
\end{equation}

We also need an additional differential equation for $\delta\varphi$ which is given by \eqref{eq:dphibar}:
\begin{equation}\label{eq:A_dphi}
    k^2 \delta \varphi + 3\varphi'' = 3(\Phi' + \Psi')
\end{equation}

\subsection{Initial conditions for the coupled ODEs} 
        
In the early universe for small $k$, the first equation in \eqref{eq:A_einstein} reduces to:
\begin{equation}\label{eq:intermediateBCeq}
    \begin{split}
        & \frac{\Psi'}{\eta}+ \frac{\Phi}{\eta^2}- \frac{2\Tilde{\alpha}_C}{(\Omega_{r0}+\Tilde{\alpha}_C)\eta^2}(\Phi - \delta\varphi') = -2\frac{8 \pi G a^2 \rho}{3}(\frac{\rho_{\gamma 0}}{\rho}\Theta_{0} + \frac{\rho_{\nu 0}}{\rho} \mathcal{N}_0) \implies \\
        & \frac{\Psi'}{\eta}+ \frac{\Phi}{\eta^2}- \frac{2\Tilde{\alpha}_C}{(\Omega_{r0}+\Tilde{\alpha}_C)\eta^2}(\Phi - \delta\varphi') = - \frac{2}{\eta^2}(\frac{\Omega_{r0}}{\Omega_{r0} + \Tilde{\alpha}_C})((1-f_{\nu})\Theta_{0} + f_{\nu} \mathcal{N}_0)
    \end{split}
\end{equation}
where we have used the background equation $ \frac{8\pi G a^2 \rho}{3} = \mathcal{H}^2 - \frac{H_0^2\Tilde{\alpha}_C}{a^2}$, so that $\mathcal{H}=\frac{1}{\eta}$ and $a=H_0\sqrt{\Omega_{r0}+\Tilde{\alpha}_C}\eta$, and where $f_{\nu}=\frac{\rho_{\nu}}{\rho_r + \rho_{\nu}}= \frac{1}{\frac{8}{7N_{\textrm{eff}}}(\frac{11}{4})^{4/3}+1}$ when all neutrinos are massless. Now we use the fact that for small $k$, $\Psi' = \Theta_{ 0}' = \mathcal{N}_0'$ together with the derivative of \eqref{eq:intermediateBCeq}$\times \eta^2$ to get:

\begin{equation}
    \eta \Psi'' + \Psi' + (1 - \frac{2\Tilde{\alpha}_C}{\Tilde{\alpha}_C + \Omega_{r0}})\Phi' + \frac{2\Tilde{\alpha}_C}{\Omega_{r0}+\Tilde{\alpha}_C}\delta \varphi'' = -\frac{2\Omega_{r0}}{\Omega_{r0} + \Tilde{\alpha}_C}\Psi'
\end{equation}

Next we employ the ansatz $\Psi = (1+\epsilon)\Phi$ in the early universe, where $\epsilon$ is a constant to be determined. Then, using \eqref{eq:A_einstein} for small $k$ (i.e. $\delta \varphi'' = (2+\epsilon)\Phi'$), we get:

\begin{equation}
    \begin{split}
        &(1+\epsilon)\eta\Phi'' + \left(2+\epsilon - \frac{2\Tilde{\alpha}_C}{\Omega_{r0}+\Tilde{\alpha}_C}+\frac{2\Omega_{r0}}{\Omega_{r0}+\Tilde{\alpha}_C}(1+\epsilon) + \frac{2\Tilde{\alpha}_C}{\Omega_{r0}+\Tilde{\alpha}_C}(2+\epsilon)\right) \Phi' = 0 \implies \\
        & (1+\epsilon) \eta \Phi'' +(4+3\epsilon)\Phi' = 0
    \end{split}
\end{equation}

This equation is solved by $\Phi = \eta^P$ for $P = 0, 1-\frac{4+3\epsilon}{1+\epsilon}$, which are respectively a growing and a decaying mode for $\epsilon > -1$.This means the early-universe growing mode has solution $\Phi = \Phi_i$ (defined at some early time $\eta_i$) which is a constant. This implies $\delta \varphi' = \delta \varphi'_i$ is also a constant in the early universe (at some early time $\eta_i$).

We can now find the initial conditions for other quantities as a function of $\Phi_i$ and $\delta \varphi'_i$ both. We start with the bottom line of equation \eqref{eq:intermediateBCeq} with $\Psi' = 0$ and $\Theta_{0,i}=\mathcal{N}_{0,i}$ to get:
\begin{equation}
    \begin{split}
        &(1-\frac{2\Tilde{\alpha}_C}{\Omega_{r0} + \Tilde{\alpha}_C})\Phi_i + \frac{2\Tilde{\alpha}_C}{\Omega_{r0} + \Tilde{\alpha}_C} \delta \varphi'_i = -\frac{2\Omega_{r0}}{\Omega_{r,0} + \Tilde{\alpha}_C}\Theta_{0,i} \implies \\
        &\boxed{\Theta_{0,i} = -\frac{1}{2}\Big\{ \frac{\Omega_{r0} -\Tilde{\alpha}_C}{\Omega_{r0}}\Phi_i + \frac{2\Tilde{\alpha}_C}{\Omega_{r0}}\delta \varphi'_i \Big\}}
    \end{split}
\end{equation}

Next, we use the third equation in \eqref{eq:A_boltzmann} as well as the multipole expansion of the first equation in \eqref{eq:A_boltzmann} (see the first equation in (22) from \cite{callin2006calculate}) with adiabatic initial conditions ($\frac{\delta}{\Theta_{0}}$ is constant) to get:
\begin{equation}
    \boxed{\delta_i = 3\Theta_{0,i}}
\end{equation}

We also use the first equation in (22) from \cite{callin2006calculate} with $\Phi' = 0, 3\Theta_1 + v_b \approx 0$ and $\Theta_2 \ll 1$ to get:
\begin{equation}
    \Theta_{1,i} = \frac{k\eta}{3}(\Phi_i + \Theta_{0,i}) \implies \boxed{\Theta_{1,i} = \frac{k\Phi_i}{6\mathcal{H}}\Big\{ \frac{
    \Omega_{r0} + \Tilde{\alpha}_C}{\Omega_{r0}} - \frac{2\Tilde{\alpha}_C}{\Omega_{r0}}\frac{\delta \varphi'_i}{\Phi_i}\Big\}}
\end{equation}

The third equation in \eqref{eq:A_boltzmann} as well as the first equation in (22) from \cite{callin2006calculate} (with adiabatic initial conditions) also give:
\begin{equation}
    -k^2v_i = (3\Phi'-\delta')_i = (3\Theta_{0}' + 3k\Theta_{1} - \delta')_i \approx 3k\Theta_{1,i} \implies   \boxed{\Theta_{1,i} = - \frac{kv_i}{3}}
\end{equation}

To find the value of $\epsilon$, we first note that the seventh and eighth equations in (22) from \cite{callin2006calculate} in the early universe imply:
        \begin{equation}\label{eq:A_nu1}
            \mathcal{N}_2'' \approx \frac{2k}{5}\mathcal{N}_1' = \frac{2k^2}{15}(\mathcal{N}_0-2\MC{N}_2 + \Phi)
        \end{equation}
        and the second equation in \eqref{eq:A_einstein} with $\Theta_2 << \mathcal{N}_2$ and $a =H_0 \sqrt{\Omega_{r0}+\Tilde{\alpha}_C}\eta$ implies:
        \begin{equation}\label{eq:A_nu2}
            \mathcal{N}_2 = -\frac{\Omega_{r0} +\Tilde{\alpha}_C}{\Omega_{r0}}\frac{(k\eta)^2(\Phi-\Psi)}{12f_{\nu}}
        \end{equation}

        Differentiating \eqref{eq:A_nu2} and equating to \eqref{eq:A_nu1} (with $\mathcal{N}_2 \ll |\Phi +\mathcal{N}_0|$) we get:
        \begin{equation}
            \mathcal{N}_0 = \frac{5}{4f_{\nu}} \left(\frac{\Omega_{r0} +\Tilde{\alpha}_C}{\Omega_{r0}}\right)\Psi - \left(\frac{5}{4f_{\nu}} \left(\frac{\Omega_{r0} +\Tilde{\alpha}_C}{\Omega_{r0}}\right)+1\right)\Phi
        \end{equation}
        which, using $\mathcal{N}_{0,i} = \Theta_{0,i} = -\frac{1}{2}\Big\{ \frac{\Omega_{r0} -\Tilde{\alpha}_C}{\Omega_{r0}}\Phi_i + \frac{2\Tilde{\alpha}_C}{\Omega_{r0}}\delta \varphi'_i \Big\}$ implies:
        \begin{equation}
            \begin{split}
                & \Psi_i = \frac{4f_{\nu}}{5}\left(\frac{\Omega_{r0}}{\Omega_{r0}+\Tilde{\alpha}_C}\right)\left\{\left(1 - \frac{\Omega_{r0} -\Tilde{\alpha}_C}{2\Omega_{r0}} + \frac{5}{4f_{\nu}} \left(\frac{\Omega_{r0} +\Tilde{\alpha}_C}{\Omega_{r0}}\right) \right) \Phi_i -\frac{\Tilde{\alpha}_C}{\Omega_{r0}}\delta\varphi'_i\right\} \implies \\
                &\boxed{\Psi_i = \left(\frac{2f_{\nu}}{5}+1\right) \Phi_i -\frac{4f_{\nu}\Tilde{\alpha}_C}{5(\Omega_{r0}+\Tilde{\alpha}_C)}\delta\varphi'_i}
            \end{split}
        \end{equation}
This satisfies the requirements of the ansatz for sufficiently small $\Tilde{\alpha}_C$, $f_{\nu}$ and $\delta\varphi'_i$. It also allows us to find:
        \begin{equation}
            \boxed{\mathcal{N}_{2,i} = \frac{k^2a^2}{12H_0^2\Omega_{\nu0}\left(\frac{5}{2f_{\nu}}+1\right)}\left(\Psi_i + \frac{2\Tilde{\alpha}_C}{\Omega_{r0}+\Tilde{\alpha}_C}\delta\varphi'_i\right)}, \boxed{\mathcal{N}_{\ell} = \frac{k}{(2\ell + 1) \mathcal{H}}\mathcal{N}_{\ell -1}, \ell > 2 }
        \end{equation}
        The derivation for the moments of the photon polarization perturbations can be found in \cite{callin2006calculate}. Turning now to the scalar field, we notice that in the early universe \eqref{eq:A_dphi} is the equation for a simple harmonic oscillator with frequency $\omega = \frac{k}{\sqrt{3}}$. This will have a general homogeneous solution:
        \begin{equation}
            \delta \varphi = \delta \varphi_i \textrm{cos}\left(\frac{k}{\sqrt{3}}\eta\right) + \frac{\sqrt{3}\delta \varphi'_i}{k}\textrm{sin}\left(\frac{k}{\sqrt{3}}\eta\right)
        \end{equation}
        where $\delta \varphi_i$ is the initial value of the scalar field at some early time $\eta_i$ and $\delta \varphi_i'$ is the initial value of its derivative.
        When $\Phi' \neq 0$, there will also be a particular integral solution to equation \eqref{eq:A_dphi}. In the early universe, if $\Phi' \lesssim \mathcal{O}(\eta)$, this particular solution will be $\delta \varphi_i \lesssim \mathcal{O}(\eta^3) \implies \left(\frac{\textrm{d}\delta\varphi}{\textrm{d}(\textrm{ln}a)}\right)_i = \mathcal{O}(\eta_i^3) \approx 0$.

        In the interest of simplicity we look at the special initial condition in which the scalar field is initially at rest at a minimum of its effective potential, with $\delta \varphi_i = \delta \varphi_i' = 0$.  However, we acknowledge that in general this theory is not necessarily constructed to behave like GR in the early universe. Therefore an interesting avenue for future research could be the exploration of the effect of different initial conditions on perturbations in 4DEGB.

We can then write down the relevant initial conditions as:
\begin{equation}
    \begin{split}
        & \Theta_{0,i} = -\frac{1}{2}\left(\frac{\Omega_{r0} - \Tilde{\alpha}_C}{\Omega_{r0}}\Phi_i\right) \\
        & \delta_{m,i} = 3\Theta_{0,i} \\
        & \Theta_{1,i} = \frac{k\Phi_i}{6\mathcal{H}(\eta_i)}\left(\frac{\Omega_{r0} + \Tilde{\alpha}_C}{\Omega_{r0}}\Phi_i\right) \\
        & v_{m,i} = -\frac{3}{k}\Theta_{1,i} \\
        & \mathcal{N}_{0,i} = \Theta_{0,i} \\
        & \left(\frac{\textrm{d}\delta\varphi}{\textrm{d}(\textrm{ln}a)}\right)_i = \mathcal{O}(\eta_i^3) \approx 0 \\
        & \delta\varphi_i = \frac{1}{3}\left(\frac{\textrm{d}\delta\varphi}{\textrm{d}(\textrm{ln}a)}\right)_i \approx 0 
    \end{split}
\end{equation}
and all other initial conditions have the same form as their GR equivalent as described in \cite{callin2006calculate}.

\bibliographystyle{unsrtnat}
\bibliography{references}

\end{document}